\pdfoutput=1

\documentclass[sigconf]{acmart}
\setcopyright{none}
\usepackage{tcolorbox}
\newcommand{\ignore}[1]{}
\usepackage{fancyhdr}
\usepackage{times}
\usepackage[normalem]{ulem}
 \usepackage[para,flushleft]{threeparttable}
\usepackage{amsfonts}

\renewcommand\footnotetextcopyrightpermission[1]{} 

\title{\vspace{0.15in}A Case for Superconducting Accelerators \vspace{0.25in}}

\begin{document}

\author{Swamit S. Tannu}
\affiliation{
  \institution{Georgia Institute of Technology}
}
\email{swamit@gatech.edu}

\author{Poulami Das}
\affiliation{
  \institution{Georgia Institute of Technology}
}

\author{Michael L. Lewis}
\affiliation{
  \institution{Microsoft}
}

\author{Robert Krick}
\affiliation{
  \institution{Microsoft}
}

\author{Douglas M. Carmean}
\affiliation{
  \institution{Microsoft}
}

\author{Moinuddin K. Qureshi}
\affiliation{
  \institution{Georgia Institute of Technology}
}
\email{moin@ece.gatech.edu}

\begin{abstract}

As the scaling of conventional CMOS-based technologies slows down, there is growing interest in alternative technologies that can improve performance or energy-efficiency.  Superconducting circuits based on Josephson Junction (JJ) is an emerging technology that can provide devices which can be switched with pico-second latencies and consuming two orders of magnitude lower switching energy compared to CMOS. While JJ-based circuits can provide high operating frequency and energy-efficiency, this technology faces three critical challenges: limited device density and lack of area-efficient technology for memory structures, reduced gate fanout compared to CMOS, and new failure modes of {\em Flux-Traps} that occurs due to the operating environment.  

The lack of dense memory technology restricts the use of superconducting technology in the near term to application domains that have high compute intensity but require negligible amount of memory.  
In this paper, we study the use of superconducting technology to build an accelerator for SHA-256 engines commonly used in Bitcoin mining applications. We show that merely porting existing CMOS-based accelerator to superconducting technology provides 10.6X improvement in energy efficiency. Redesigning the accelerator to suit the unique constraints of superconducting technology (such as low fanout) improves the energy efficiency to 12.2X. We also investigate solutions to make the accelerator tolerant of new fault modes and show how this fault-tolerant design can be leveraged to reduce the operating current, thereby increasing the overall energy-efficiency to 46X compared to CMOS. Our paper also develops a workflow for evaluating area, performance, and power for accelerators built in superconducting technology, and this workflow can help other researchers explore designs using this technology. 
\end{abstract}

\maketitle
\thispagestyle{firstpage}
\pagestyle{plain}


\section{Introduction}
  

Slowdown in Moore's law limits the energy-efficiency and performance that can be obtained with general purpose computers. To bridge the gap between available performance and application demand, system designs are increasingly moving towards building application-specific accelerators~\cite{TPU,brainwave}. While accelerators provide significant performance and energy-efficiency gains, the continued performance growth offered by accelerators also gets affected by technology scaling. Unfortunately, the marginal improvements in CMOS device density and performance forces us to investigate alternative technologies that can provide improved performance and energy-efficiency. Superconducting technology is one such potential candidate. However, the technology has several constraints and it is not yet mature to support complex designs. This paper presents a case for accelerators based on emerging superconducting technology.

\vspace{0.05 in}
\noindent{\bf What is the Technology?} Superconductivity is a physical phenomenon observed in certain metals that exhibit zero electric resistance at extremely low temperatures. It can be leveraged to build energy efficient and high-performance switching devices known as Josephson Junctions (JJ). JJ devices are used as the basic building blocks for constructing logic and memory circuits. JJ technology can operate at frequencies up to 20 GHz, due to minimal switching delay (2 pico-seconds) of JJs and lossless wires. Moreover, the switching energy of a JJ is about five orders of magnitudes smaller than CMOS. However, to achieve superconductivity, JJ devices need to be operated at temperatures close to few Kelvins (typically 4K). To maintain such temperature, a cryogenic cooler is used, and such coolers typically consume 300W power for every 1W dissipated at 4K. Although cooling overhead seems significant, the low switching energy of JJ can still enable devices that have 100x lower energy consumption over CMOS even after accounting for cooling energy overhead~\cite{herr2011ultra}. Thus, JJ technology can provide significant improvements in operating frequency and energy-efficiency compared to CMOS.


\begin{figure}[htb]
\centering
    \includegraphics[width=0.95\columnwidth]{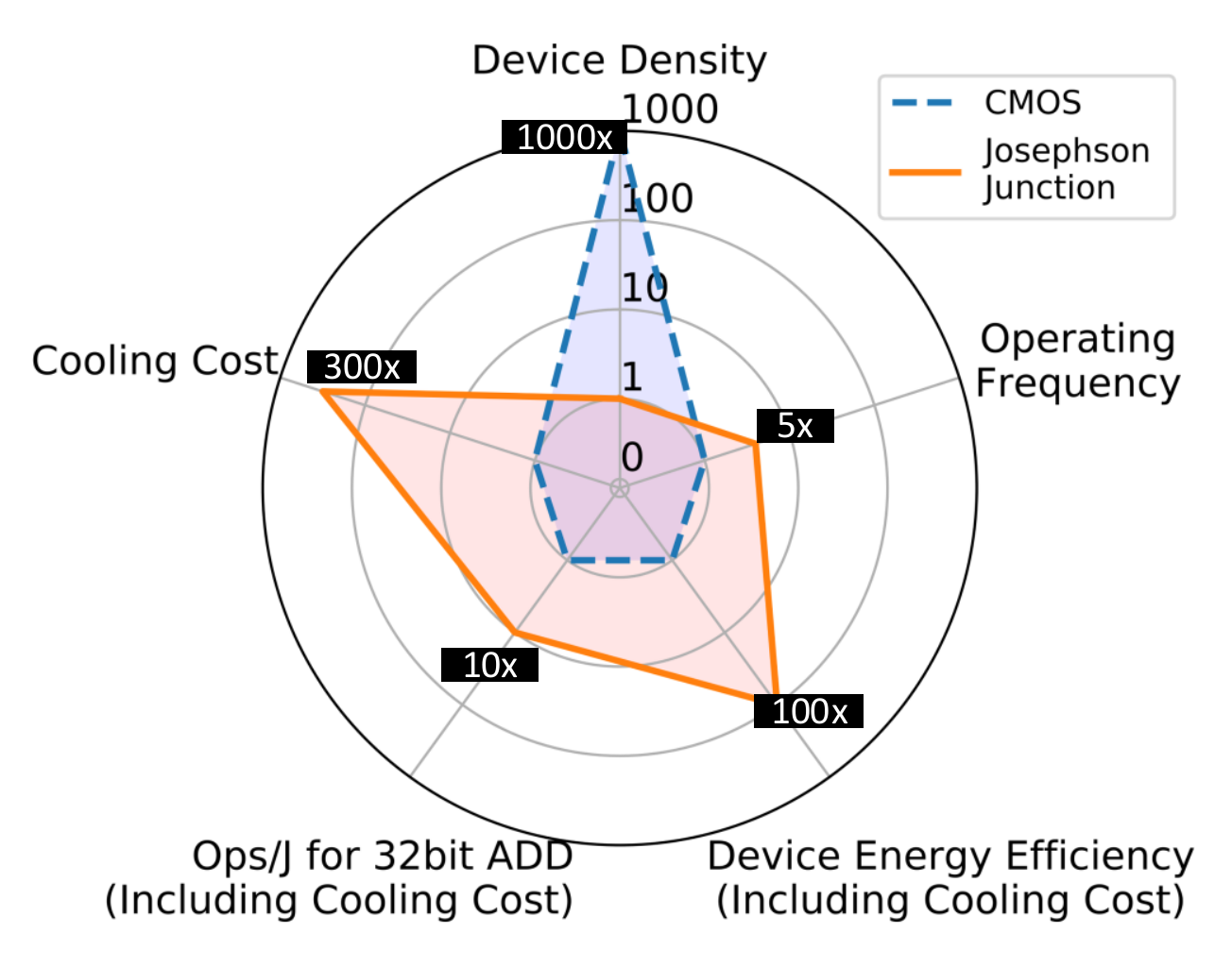}
    \vspace{-0.15in}
    \caption{Comparing Josephson Junction (JJ) with 16-nm CMOS, based on parameters derived from~\cite{dorojevets2015towards,pedram,holmes2013energy}}
    \label{fig:intro}
    \vspace{-0.15in}
\end{figure}

\vspace{0.05 in}
\noindent{\bf What are the Challenges?}  Building a JJ based computing system is challenging. The primary challenge is limited logic and memory density of JJ-technology. For existing process technology with 248nm feature size, JJ-device density lags by 1000x as compared to CMOS~\cite{mit_ll,tolpygo2016superconductor}. Although JJ density is projected to grow in coming years~\cite{holmes2013energy}, near-term JJ-technology may not be able to close the 1000x density gap between CMOS and JJ devices. JJ based logic requires more devices per gate as compared to CMOS. The higher device complexity results from limited output driving capacity of JJs. For example, standard CMOS gates have fan-out of four, whereas, JJ based logic gates can drive at most one output without requiring extra output drivers. These output drivers are known as Josephson transmission lines (JTLs) and costs 2-JJ devices per JTL exacerbating the density problem. The limited fan-out results in significantly different design trade-offs for accelerators built in superconducting technology compared to conventional CMOS. The third challenge is the reliability of JJ devices that is susceptible to magnetic flux-trapping and manufacturing defects. These defects can result in intermittent faults. 
In this paper, we study near-term JJ technology for building accelerators and make the following contributions.


\vspace{0.05 in}
\noindent
\textbf{Contribution-1: Study of Superconducting Accelerator:} Given the lack of dense memory technology,  accelerators built with JJ based technology are likely to be restricted to applications with tiny working set size and high computational intensity, to obtain significant performance and energy efficiency improvements over baseline CMOS accelerators. In this paper, we focus on a SHA-256 accelerator used for block-chain applications or bitcoin mining. We choose a SHA accelerator due to its simple yet rich design that enables unique design trade-offs offered by JJ technology. Bitcoin mining application requires repeated computation of the double SHA-256 hash for an input message and a 32-bit random key known as a \texttt{nonce}. A bitcoin miner repeats the SHA computation until it finds a key that produces a hash with a set number of leading zeros. This repeated evaluation with the same input message but with different keys fit well with the JJ constraints as the compute intensity is exceptionally high with tiny memory footprint. Also, blockchain applications have a concrete figure-of-merit for both performance (Giga-hashes per Second, or GH/S) and energy-efficiency (Giga-hashes per Joule, or GH/J). Furthermore, existing CMOS bitcoin ASICs serves as highly optimized baseline facilitating  technology to technology comparison to evaluate the system-level benefits of JJ based fixed function accelerators. We use Goldstrike 1 ~\cite{Goldstrike} bitcoin miner as the baseline CMOS design.\footnote{Bitcoing mining is a competitive industry and the state-of-the-art industrial designs of bitcoin mining accelerators are often kept proprietary by the companies.  We use Goldstrike 1 in our evalutions because the implemenation is publicly available and we can make a fair comparison of CMOS versus JJ technology.  Given that the design details for state-of-the-art bitcoin mining accelerators, such as Antminer S-9 (16nm ASIC~\cite{debitcoin}), are not publicly available, we are unable to evaluate such designs in JJ technology. Nonetheless, we do compare the energy-efficiency of our proposal with the publicly reported energy-efficiency number of Antminer S-9 in Section~\ref{sec:evaluation} and observe a 15x improvement (including the cooling overheads).}


\vspace{0.05 in}
\noindent
\textbf{Contribution-2: Technology-Aware Design:} JJ-based adders have significantly different area and performance trade-offs compared to CMOS designs. For example, by accounting for the fan-out problem, we choose an adder structure that minimizes the overall JTL count by fusing consecutive additions. Similarly,  the baseline design requires per-stage registers to store the temporary values and relies on wide buses, both of which incur significant overheads. To reduce both JTL and memory overhead, we leverage a predictable register production and consumption of intermediate  variables. Instead of storing the intermediate variables in register files we use extremely resource and energy efficient delay-lines to synchronize the producer and consumer stages. Redesigning the accelerator to  JJ-technology specific constraints improve the performance by 1.8x and increases the energy-efficiency from 10.6x to 12.4x compared to CMOS implementation.

\vspace{0.05 in}
\noindent
\textbf{Contribution-3: Fault Mitigation and BTWC:}  Superconducting technology has a significantly different fault mode. The source of the fault lies in the operating condition and fundamental property of superconductivity known as Flux trapping. It results in correlated and large granularity faults. Furthermore these faults have relatively longer life time i.e they manifest for longer than transients but they are not permanent. To the best of our knowledge, this is the first paper to mitigate faults in JJ technology by leveraging architecture-level solutions, such as redundancy and sparing.  To improve the reliability of the proposed accelerator, we design a fault-tolerant SHA-256 engine by provisioning one additional (spare) pipeline stage and a bypassing mechanism that can detect and protect the accelerator against large granularity faults.  Our fault-tolerant design incurs minor storage overhead; however, it can be leveraged to improve energy-efficiency. For example, in superconducting circuits power is a product of critical current ($I_c$) and operating frequency.  The critical current  is essential for the correct operation of a circuit. Minimum value of $I_c$ ensures a noise margin dictating tolerable error rate in the logic and memory circuits. This trade-off between $I_c$ and error rate can be leveraged to operate the accelerator at {\em Better-Than-Worst-Case (BTWC)} operating point by leveraging the fault tolerance circuitry.  Such, BTWC design can reduce the $I_c$ from 38 to 10 micro-amperes, improving the overall energy-efficiency of the superconducting accelerator to 46x compared to the CMOS implementation.

\vspace{0.05 in}
\noindent
\textbf{Contribution-4: Methodology for Estimating Area, Performance and Power of Superconducting Accelerators:} Estimation of performance, power, and area for superconducting logic is difficult due to lack of automated design tools. Furthermore, standard cells and design rules in superconducting logic families are fundamentally different from CMOS technology. For example, logic cells in JJ technology has limited driving capacity, and to drive more than one cell, a buffer like cell known as Josephson Junction Transmission Line (JTLs) must be placed between two cells. This limits the direct usage of standard CMOS tools to perform a design space exploration for superconducting technology. To overcome this problem, we use open-source back-end design tools to incorporate design constraints specific to superconducting logic. In addition to the modified design tool, we use analytical models to calculate performance, power, and area for superconducting logic. We introduce a workflow and methodology to explore design space for accelerators built in superconducting logic. Such a workflow can help other researchers in exploring different designs at the architecture level using  this emerging technology.

\ignore{

The abstraction of accelrator can enable

However, significant number of prior work has demonstrated prototype interconnect technologies that can be used to connect the host and accelerator across thermal boundary without leaking excessive heat. The demonstrated bandwidth for such interconnect is reported as 30 Gbps with Bit Error Rate (BER) of $10^{-10}$. In case of long latency between 300K and 4K,  a buffer or cache at the intermediate cryogenic temperature can improve the average latency. For example, 77K (liquid nitrogen cooled) stage can be used to place a large capacity buffer that leverages conventional CMOS memory. The idea of intermediate cryogenic memory is proposed by several prior works and in fact recent industry and academic studies show the operational comodity DRAM at 77K~\cite{HTMT,tannu2017cryogenic,RAMBUS,RAMBUS1}.

However, the extreme energy efficiency can still be leveraged by distributing memory and compute across thermal hierarchy as shown in the figure~\cite{}. This  thermally distributed systems enable new types of designs trade-offs. For example, table~\ref{fig:1} shows an organization of the superconducting processor with the power, performance and area constraints.

The block chain is the foundation for the crypto-currencies, smart contacts, and many emerging applications. It relies on the on the repeated computation of the cryptography hash functions that are non-invertable. For example, bitcoin is one of the crypto-currencies that leverages the block chain technology and relies on computation of the SHA-256 hash. The bitcoin relies on the distributed network consists of individuals who verify the integrity of each transactions by performing computationally intensive tasks, and in return they receive reward in form of bitcoins by the network. This process of verifying the transaction is known as mining. Mining requires substantial energy as task of verifying the transactions rests on repeated evaluation of SHA-256 hash. A miner profits by improving energy efficiency. Bitcoin mining has evolved from using commodity CPUs and GPUs to custom ASICs ~\cite{magaki2016asic, taylor2017evolution}. Existing bitcoin ASIC designs leverage 16nm custom ASIC CMOS fab to optimize the performance and the efficiency. The difference between ASICs and CPU based mining is about 1000x in hash rate and the efficiency.

The idea of using Josephson junction for computing is around for last 30 years~\cite{clark1967} and previous research programs have demonstrated prototype systems~\cite{HTMT,IBM}. However, due to scalable competing CMOS technology and fabrication challenges the superconducting fabrication technology faced with stagnant growth. Fortunately, recent advances in the superconducting circuits and slowdown in the CMOS scaling has provided an impetus towards developing scalable fabrication technology for superconducting logic. The IARPA C-3 program is geared towards developing JJ logic to enable post-exascale solutions for HPC and data-centers~\cite{manheimer2015cryogenic}. In this paper, we highlight opportunities and challenges in building computing systems with superconducting technology.

}

\ignore{

Josephson junction is a superconductor-insulator-superconductor junction that is fabricated using superconducting (SC) fabrication process














}





\section{Superconducting Technology}

\subsection{Josephson Junction Device}
Few metals exhibit zero resistance to the flow of current at cryogenic temperatures, a phenomenon known as \textit{superconductivity}. Superconductivity can be achieved by cooling metal wires below their critical temperature. For example, Aluminum (Al) and Niobium (Nb) superconduct below 1.2K and 9.3K respectively. Superconductivity is leveraged in building a switching device called a Josephson Junction (JJ). 

\begin{figure}[!htb]
    \includegraphics[width=1.0\columnwidth]{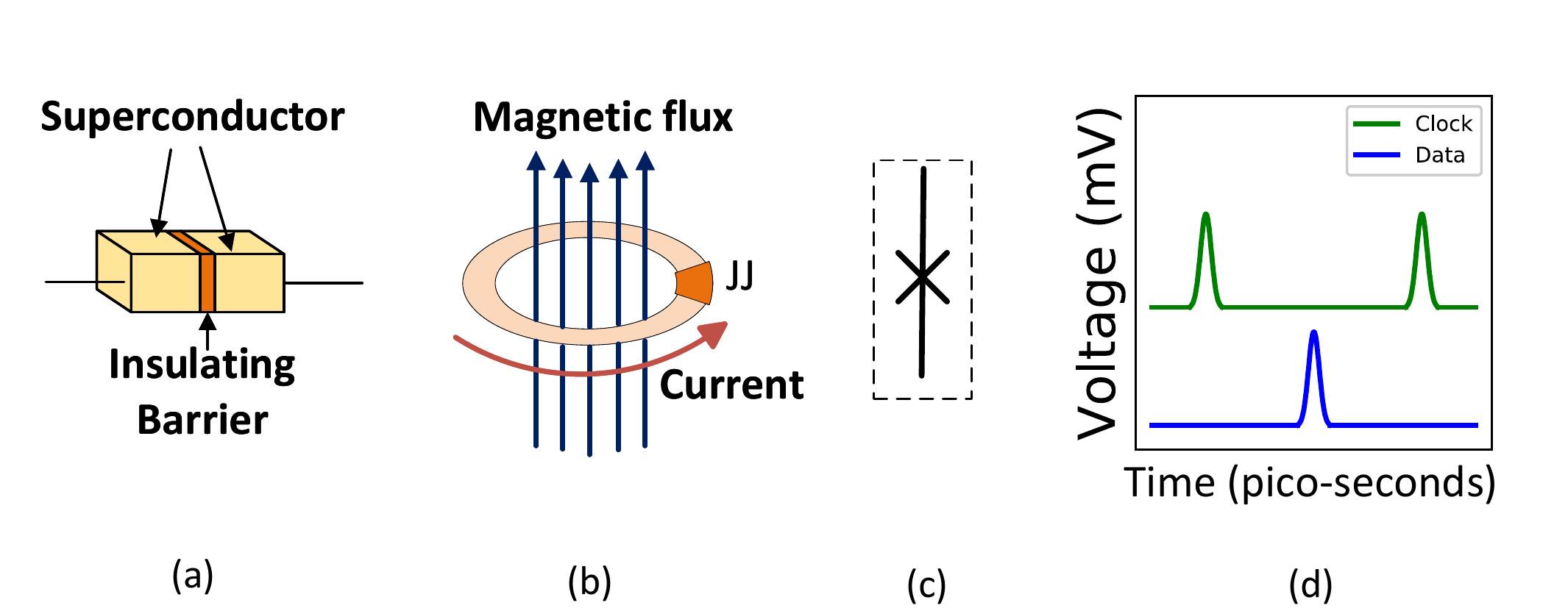}
    \vspace{-0.10in}
    \caption{(a) Josephson junction device (JJ) (b) shunted JJ (c) circuit symbol for JJ (d) Voltage-time curve.}
    
    \label{fig:jj_schema}
   \vspace{-0.05in}
\end{figure}

A JJ is fabricated by interposing a thin barrier between two superconducting wires as shown in Figure~\ref{fig:jj_schema}(a). This barrier allows the electrons to tunnel through it even in the absence of an applied voltage. Moreover, the tunneling is controlled by changing the input current. For example, when the current flowing through the device exceeds its critical current ($I_c$), a JJ switches from superconducting to a resistive state. Alternately, it goes back to the superconducting state if the current is reduced below $I_c$. 
Note that the voltage across the JJ in a superconducting state becomes zero.

\subsection{Josephson Junction as Switch}


In a superconducting loop with a JJ, the magnetic flux($\phi$) is quantized i.e. it can take only integer multiple values of a single flux quanta (SFQ) ($\phi_0$). Magnetic flux ($\phi$) is the magnetic field per unit area. SFQ is the magnetic flux generated by the tunneling of an electron pair. By switching the JJ between superconducting and resistive states, the amount of flux in the circuit can be controlled. Presence or absence of SFQ can thus be used to represent digital information ``1" and ``0" respectively. When a JJ switches from superconducting to a resistive state, the magnetic flux through the superconducting loop containing the JJ changes by a flux quanta, generating an SFQ pulse of about 1 pico-second duration and 2 milli-volt magnitude, as shown in Figure~\ref{fig:jj_schema}(d).


In superconducting technology, SFQ pulses facilitate encoding, processing, and transmission of digital information. JJs are almost ideal digital switches that are characterized by two basic properties: high-speed switching and ultra-low power dissipation. SFQ pulses can be as narrow as one pico-second making it possible to clock circuits at very high frequencies. Superconducting passive transmission lines (PTL) are also able to transmit SFQ pulses with extremely low losses at 4K. These lossless interconnects and low switching energy for JJs (2x$10^{-20}$ J) enable very low power dissipation.

\subsection{Superconducting Logic Gates Using RQL}
\label{rql}

\ignore{
Early superconducting circuits used  DC biased JJs with bias resistors to distribute power. These resistors contributed to sizeable static power dissipation limiting the scalability of resistor biased SFQ (RSFQ) circuits.  Recently proposed, Reciprocal Quantum Logic (RQL) overcomes the drawback of static power dissipation of RSFQ~\cite{herr2011ultra}. It uses AC power lines and inductive coupling to distribute power and clock to the devices in the circuit. It also uses lower device counts per gate as compared to RSFQ. These factors  enable negligible static power dissipation and higher logic-gate density. 

\begin{figure}[!htb]
 \vspace{0.1in}
    \includegraphics[width=1.0\columnwidth]{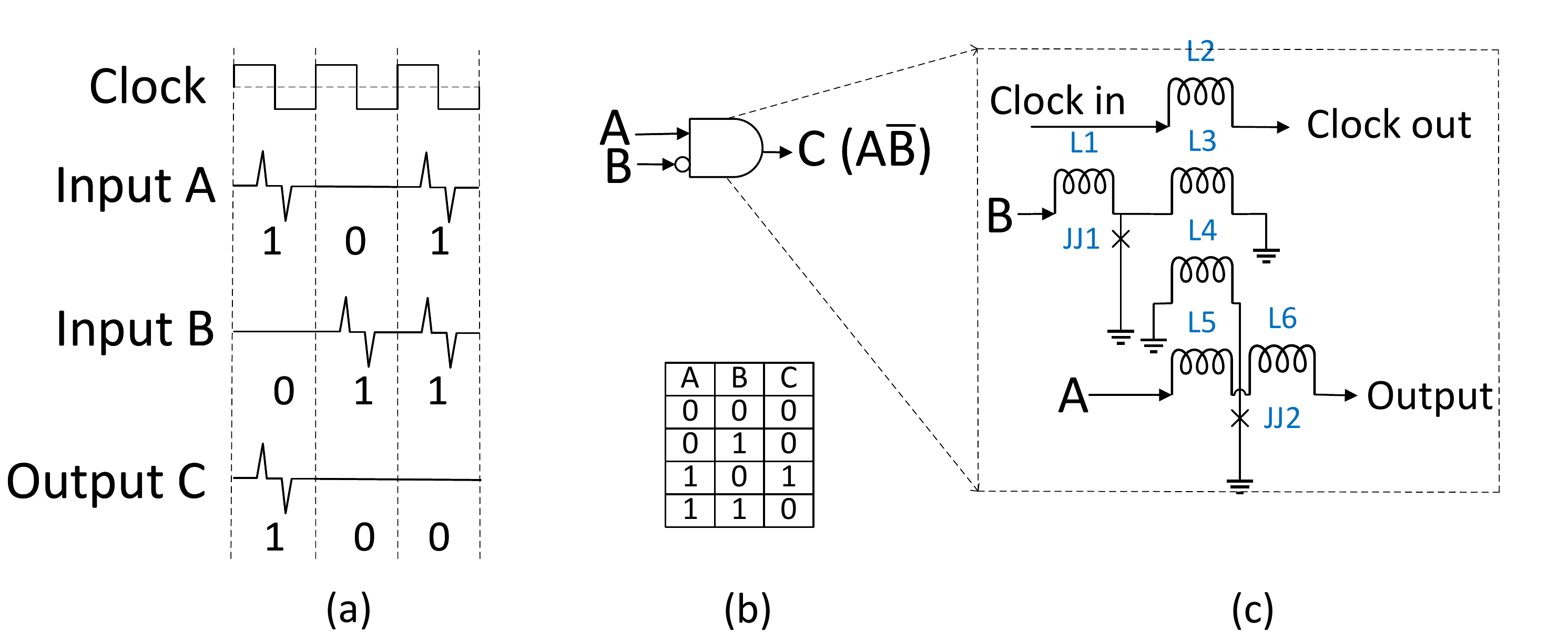}
    \caption{a) Data encoding in RQL and A-AND-NOT-B logical operation (b)Gate symbol (c) Circuit schematic}
    \label{fig:rql_basics}
     \vspace{-0.15in}
\end{figure}
}

Reciprocal Quantum Logic (RQL) uses JJ based switches such that a digital ``1" is encoded as a pair of SFQ pulses of opposite polarity and a ``0" is encoded as the absence of SFQ pulses as shown in Figure~\ref{fig:rql_basics}(a). For details of RQL logic gate design please refer to~\cite{herr2011ultra}. The RQL family consists of two universal gates, the AND-OR gate and the logical A-AND-NOT-B (referred to as A-NOT-B, as shown in Figure~\ref{fig:rql_basics}) gate that enables the design of complex circuits~\cite{oberg2011superconducting}.

\ignore{
For a digital ``1", operations with the positive SFQ pulse involve storage and routing of SFQ data. When only the positive pulses are considered, the gates act like state machines, as the input changes, the internal flux state of the inductive loops changes as well. The negative pulse resets the internal state of the RQL cells every clock cycle to the initial state and produces combinational logic behavior. The erasing operation facilitated by the negative pulse simplifies the logic design, to the extent that some gates can comprise of as low as only two active devices with biasing inductors.
}

\begin{figure}[!htb]
 \vspace{-0.05in}
    \includegraphics[width=1.0\columnwidth]{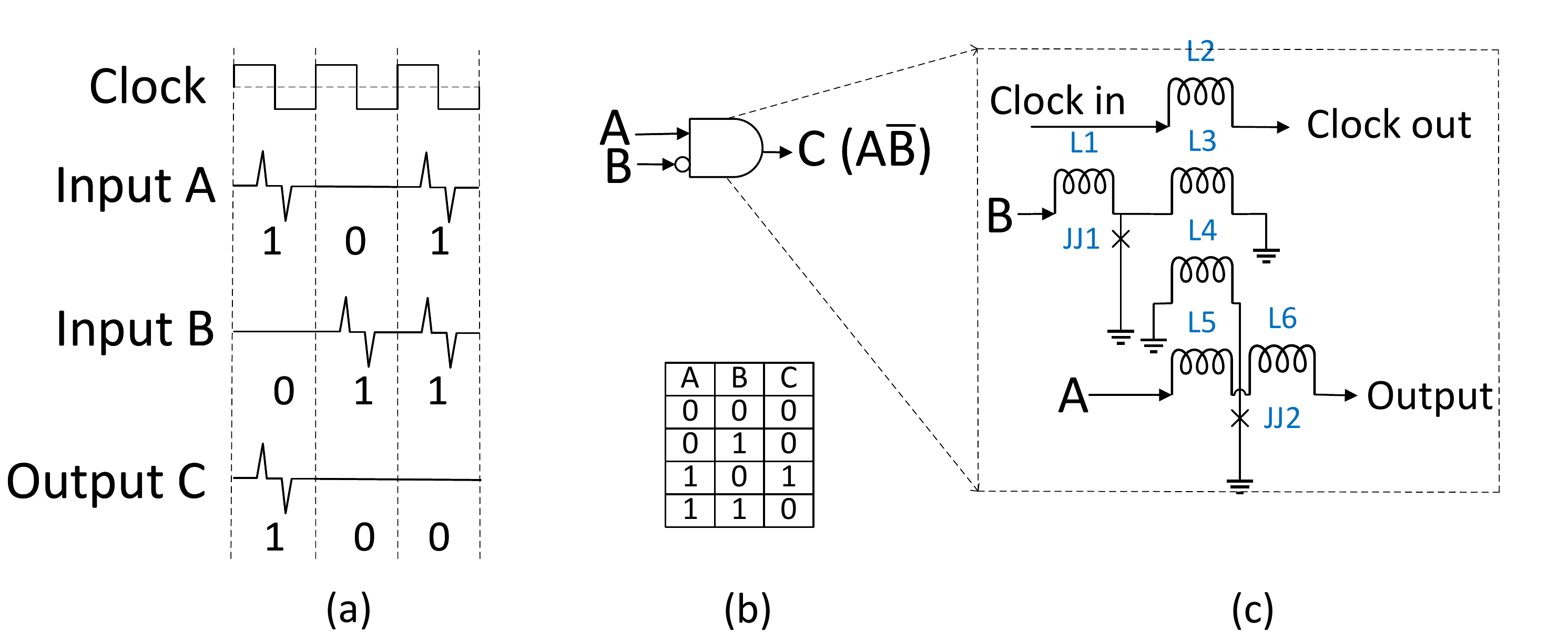}
     \vspace{-0.15in}
    \caption{a) Data encoding in RQL and A-AND-NOT-B logical operation (b) Logic Table (c) Circuit schematic}
        \vspace{-0.1in}
    \label{fig:rql_basics}
\end{figure}

Since its introduction in 2011, RQL circuits with 72,800 JJs have been demonstrated. Other demonstrated circuits include shift registers, small arithmetic circuits, transmission driver systems, and serial data receiver systems~\cite{herr20138, herr2015reproducible,herr2018superconducting,shauck2018reciprocal}. Design and resource estimates exist for 32-bit and 64-bit integer and floating-point arithmetic and logical units, register file, on-chip storage components, bloom filters ~\cite{dorojevets2015towards,dorojevets2015fast,dorojevets2018energy}.

\subsection{Memory Challenges for JJ Technology}

Table~\ref{tab:CMOSvsJJ} compares the energy-efficiency of typical operations in \texttt{16-nm} CMOS and superconducting technology. To report energy for \texttt{16-nm}, we use established ITRS scaling factors since we lack open source \texttt{16-nm} synthesis libraries \cite{wright2018standards,esmaeilzadeh2011dark}. We observe that memory operations are less energy efficient compared to arithmetic and logic operations for JJ technology. Furthermore, building memory takes more area in JJ technology as there is no dense memory solution like SRAM or DRAM currently available in the superconducting domain. Researchers are exploring superconducting memory solutions such as hybrid-JJ-CMOS memory, Josephson magnetic random access memory (JMRAM), but their capacity is likely to remain severely limited compared to conventional technologies. The limited device density and costly memory operations and capacity constrains the potential applications to computationally intensive applications that have small working sets. We explore the design of superconducting technology for one such application. 

\begin{table}[htb!]
\begin{center}

\label{tab:CMOSvsJJ}
\begin{small}
\vspace{-0.10 in}
\caption{Energy/op comparison of 16-nm CMOS and JJ-logic (\textit{including cooling overhead of 300x)} }
\setlength{\tabcolsep}{0.05cm} 
\renewcommand{\arraystretch}{1.2}
\label{tab:CMOSvsJJ}
\begin{tabular}{ |c|c|c|c| } 
\hline
Parameter                               & CMOS-16nm &  JJ at 4K  & Improvement           \\ \hline \hline
 64bit-Add                                    & 0.592 pJ    &           0.06 pJ & 9.86x                 \\
\hline

64bit-Multiply                          & 2.367 pJ &  0.248pJ & 9.54x      \\
\hline

64bit-RF-Load             &     0.050 pJ          &   0.05 pJ & 1x                                                                   \\

\ignore{
\hline
64bit DRAM Transfer                       & 2560 pJ              &   768pJ & 3.3x          & between  DRAM at 77K and 4K processor~\cite{RAMBUS}                                       \\
}
\hline
Off-chip Interconnect         &  300 pJ               &  8.6 fJ    & 30000x                                                             \\

\hline
300K to 4K link        &  -                &  3712 pJ/bit    &   NA                                                  \\

\hline

\end{tabular}
\label{table:CMOSvsJJ}
\end{small}
\vspace{-0.3 in}
\end{center}
\end{table}

\ignore{

\subsection{Challenges in Superconducting Logic}

\subsubsection{Device Density}
Existing JJ fabrication technology has limited device density. It lags behind CMOS by approximately 1000x. Currently, MIT Lincoln Lab, Hypress, and D-Wave Systems offer superconducting foundry services~\cite{mit_ll,holmes2017superconductor}. MIT Lincoln Lab's state of the art foundry offers JJ devices at 248nm and 8 metal layers~\cite{tolpygo2016superconductor}. The foundry is expected to reach device density of million JJs/$cm^2$ and has demonstrated functional circuits with the density of 100,000 JJ/$cm^2$~\cite{holmes2017superconductor} Furthermore, existing superconducting logic families lack dense memory solutions limiting the capacity of on-chip memories. Recent progress and investments in the superconducting process technology promise improvements in device density~\cite{manheimer2015cryogenic}.

\subsubsection{Gate Fan-out challenge}
RQL gates have limited fan-out and {\em Josephson transmission line (JTL)} is used to drive the gates. JTLs amplify the SFQ pulse energy and provide fan-out capacity similar to buffers in traditional CMOS circuits. However, JTL overhead is significantly higher as compared CMOS buffer due to an inherently limited driving capacity of RQL gates. RQL gates require JTL for every output load. Furthermore, the number of JTLs scale linearly with distance between the gates. Each JTL comprises 2 JJs. Typically, a complex logic circuit with higher fan-out requires a large number of JJ devices due to high JTL cost. \ignore{Furthermore, JTL cost can be substantially high for process nodes with limited metal layers. Fortunately, next-generation JJ processes technology enables 8 to 10 metal layers alleviating the high JTL overhead due to long routing paths. However, even with such optimizations, the overhead of JTL continues to be significant for JJ-based circuits.}

\ignore{
Most CMOS designs use a variety of gates with different input sizes and drive strengths. This level of flexibility is not yet available for RQL universal gates. RQL gates available today are only 2-input gates. This poses a challenge in logic optimizations and therefore scaling designs. The gate complexity and associated JTL overheads for complex designs, therefore, rise fairly quick as the number of variables in the functions increase.}

\subsubsection{Device and System Reliability}
\label{sec:realibility}
Similar to CMOS circuits, superconducting circuits experience manufacturing, and timing faults. Besides conventional faults and sources of faults, JJ technology experiences another unique source of errors known as operational faults resulting from a non-ideal environment. For example, when the superconducting circuit is cooled down, not all the components achieve superconductivity at the same time, this results in the trapping of the stray magnetic field, a phenomenon known as {\em Flux Trapping}. Flux trapping is a direct consequence of a superconducting property known as Meissner effect. For example, when metal achieves superconductivity, it expels all the magnetic field lines and acts as a perfect magnetic shield. However, while cooling the metal if it is not cooled uniformly, magnetic flux is trapped in parts of the chip. 

The trapped flux often known as flux vortex can result in non-functional or partially functioning superconducting circuits. Furthermore, flux-trapping reduces the noise margin significantly forcing us to operate at the reduced clock frequency. Flux trapping is a significant challenge in scaling the superconducting circuits. Fortunately, recent demonstrations show that the problem of flux trapping can be reduced by introducing motes -- an extra metal layer with holes, that can be used to attract and trap the flux vortexes away from the circuits. While such solutions can provide some protection against flux trapping, flux trapping cannot be eliminated entirely and continues to be a reliability challenge for JJ-based circuits.   

}

\ignore{
\subsection{Why Superconducting Accelerators?}

Superconducting circuits offer high energy efficiency. However, with limited device density and memory capacity, designing superconducting general purpose computers is incredibly challenging. Furthermore, lack of sophisticated design tools exacerbate the density problem as existing CMOS-based synthesis tools used for JJ design can not maximize device utilization. We believe that both of the problems are related to design and manufacturing economics rather than being fundamental challenges. However, until the technology reaches the maturity to manufacture and test billion Josephson Junctions per cm$^2$, which is typically required for general-purpose computing, we can leverage the existing fabrication process to build accelerators.
}


\ignore{
\subsubsection{System Organization}
We propose a superconducting accelerator that operates at 4K and connects to a host either using a cryogenic DRAM or interconnects as shown in the Figure~\ref{fig:org}. JJ-accelerator will require scalable cooling solutions and the interface between host and accelerator. For the Niobium technology based circuits, the critical temperature is 9.2K ($-441^{o}$F). Fortunately, availability of large-scale Liquid Helium based coolers can be leveraged to maintain 4.2K temperature. For example, Large Hadron Collider at CERN uses superconducting wires cooled at a 1.8K temperature over 3.3 Km of sector~\cite{LHC}. Cryogenic cooling at 4K is a mature technology with the cooling overheads of cryo-cooler ranging from 300 W/W to 1000W/W~\cite{holmes2013energy}. The cooler provides a  thermal guard band against local heating and thermal noise. 


\begin{figure}[!htp]

\centering
    \includegraphics[width=0.8\columnwidth]{Figure/Organization_horizontal.pdf}
    \caption{Superconducting Accelerator Organization}
\label{fig:org}    
\vspace{-0.15in}
\end{figure}

For interfacing, the host and accelerator can be either done via a direct link between host and  accelerator or via cryogenic memories. A cryogenic data-link across 300K and 4K has been demonstrated by Ravindran et al. ~\cite{ravindran2015power}. The reported interconnect design use three stage amplifiers that are distributed across thermal hierarchy to achieve 30 Gbps bandwidth and to consume $140\mu$W power at 4K. Note that superconducting circuits utilize SFQ pulses with few milli-volts as opposed to CMOS circuits that require at least 400mV. If off-chip memory is required, then superconducting accelerators can use Cryogenic DRAM~\cite{tannu2017cryogenic, RAMBUS1,RAMBUS} to avoid the significant thermal gradient between superconducting domain and room temperature, as shown in Figure~\ref{fig:org}(1).  Without loss of generality, in our paper, we use the direct-access model  for our superconducting accelerator as it requires negligible amount of memory.

Before we describe our accelerator, we discuss the methodology for evaluating the area, performance, and power for the accelerators based on JJ technology.

}


\ignore{

\begin{figure}[!htb]
    \includegraphics[width=1.0\columnwidth]{Figure/JJ_density.pdf}
    \caption{Trends in the number of JJs fabricated over the years}
    \label{fig:jj_scaling}
\end{figure}

Furthermore, superconducting memories are built using storage cells that requires to be connected through interconnection cells. This adds to a significant overhead in terms of device counts. Current fabrication technologies offer limited device density which worsens the problem of fabricating memory devices in superconducting logic families. All these factors result in low capacity. Alternate options of memory at low temperatures include  Josephson magnetic random-access memory (JMRAM) ~\cite{herr2012josephson}. This memory technology is still in its infancy and offers only a limited capacity. Recent experiments ~\cite{dayton2018experimental} demonstrated this type of memory to be scalable, dense and energy-efficient with respect to its previous counterparts. Hybrid memories that use both JJs and CMOS, provide slightly better capacity, but are not energy-efficient when operated at 4 K ~\cite{liu2007latency}.

A feasibility study in ~\cite{holmes2013energy} shows memory scaling to be of utmost priority. The feasibility of developing a fully functional superconducting general purpose computing landscape looks challenging at this moment~\cite{tolpygo2016superconductor}. A key factor behind this is the slow projected growth of the number of devices (doubles every 4.5 years), way slower than CMOS technology. Figure~\ref{fig:jj_scaling} shows the number of fabricated JJs to demonstrate working superconducting technology circuits over the last 20 years. Flux trapping is one of the primary causes hindering device scalability.

This phenomenon is stochastic and can produce reduced operational margins and non-functional circuits. The projected device density for some proposed process nodes from MIT Lincoln Laboratory are presented in Table ~\ref{tab:process_nodes}. HYPRES currently offers foundry services with critical current densities of 10, 30, 100, 4500, and 10000 kA/cm$^2$. They use photo-lithography tool based on a planarized chip fabrication process called RIPPLE ~\cite{yohannes2015planarized}. They are making continued efforts to provide superconducting chips that can feature critical current densities such as 4.5 kA/cm$^2$, 100 A/cm$^2$, 30 A/cm$^2$. D-Wave Systems is another vendor that offers foundry services for superconducting technology ~\cite{bunyk2014architectural,lanting2014entanglement}.

So far, a few designs using RQL family a few designs have been fabricated including a 200-bit shift register ~\cite{herr2011ultra} and a 8-bit carry-lookahead adder ~\cite{herr20138}. Design of some key components of a processor such as 32-bit and 64-bit integer and floating-point arithmetic and logical units, on-chip storage elements including registers, memory with read and write ports, decoders, FIFOs (first-in first-out buffers) are presented in ~\cite{dorojevets2015towards}. The design and energy-efficiency analysis of 32-bit and 64-bit storage structures such as random access memory, register files, direct mapped caches, and FIFOs is presented in ~\cite{dorojevets2015fast}. In ~\cite{holmes2013energy}, the feasibility of a computing system performing in the range of 1 to 1000 PFLOPs is analyzed.

\subsection{Cryogenic DRAM}
Superconducting logic operating at 4.2 K cannot work in conjunction with memory sitting at room temperature (300 K) for large systems. The huge thermal gradient causes the wires connecting the two thermal domains to leak power very rapidly, thereby limiting the scalability of the system. As already described in section ~\ref{rql}, memories built using RQL storage cells are currently unable to offer large capacities. JJ-MRAM is a proposed memory technology that can offer capacity in an energy-efficient manner at cryogenic temperatures. However, this type of memory is in its early stage of development as we speak offering only a few megabytes of capacity [?]. CMOS DRAM memories using long-channel planar devices were demonstrated to work at 50 K ~\cite{henkels1989low}. The authors in ~\cite{tannu2017cryogenic} looked at the feasibility of using more recent off the shelf commercially available DRAM chips at low temperatures. It was revealed that modern DRAMs, built from non-planar short channel devices, can operate at up to 80 K and occasional errors can be mitigated by conventional schemes like Chipkill or sparing. This looks attractive because cryogenic refrigerators offer multiple temperature levels, allowing us to add a layer in the thermal hierarchy to overcome memory capacity limitations.

Although RQL technology looks very promising in terms of energy-efficiency, design of dense and reliable memory that can operate at cryogenic temperatures is still one of the major challenges in deploying this technology for commercial products. The storage structures proposed in ~\cite{dorojevets2015towards, dorojevets2015fast} requires a large number of JJs and incurs significant overhead due to interconnection cells. This limits their storage capacity. A new type of memory is being developed now, called JJ-MRAM that offers high density in an energy-efficient manner ~\cite{herr2012josephson}. Researchers have also looked at the feasibility of using regular DRAM at a new thermal hierarchy (80 K) in ~\cite{tannu2017cryogenic}.

\begin{table}[!htb]
\caption{Specification of few Cryogenic Coolers}

\label{tab:coolers}
\begin{tabular}{ |l|l|l|l|l| } 
\hline
Model                            & SHI SRDK           & SHI GMJTC     & Linde                  & Linde      \\
                                 & 415D-F50   & G310SLCR     & LR70                   &   LR280                       \\
\hline
Refrigeration           &              &              &                &   \\
capacity (W)            &   1.5        &   10         &   100          &  1020  \\
\hline
Cryocooling             &              &              &                &   \\
efficiency (W/W)        &   5000       &   1280       &   450          & 395   \\
\hline
Power Budget:       &             &             &         &   \\
cooling (kW)        &  7.5        &  12.8       &  45     & 400  \\
\hline
Power Budget:        &               &               &           &   \\
maintenance (kW)    &  7.5          & 27.2          & 155       &  1600 \\
\hline
Total Power (kW)    & 15            & 40            & 200       &  2000 \\
\hline
Energy-efficiency   &               &               &           &       \\
(FLOPS/W)           & 0.7x$10^{11}$ &  2.5x$10^{11}$ &  5x$10^{11}$ & 5x$10^{11}$  \\
\hline
Cost                & 45 k\$        & 320 k\$       & 2 M\$     &  6 M\$  \\
\hline
\end{tabular}
\end{table}

Although it looks promising, superconducting technology still lags behind CMOS primarily due to two reasons. Firstly, technology limitations like limited device counts, limited logic density and memory capacity. Secondly, it lacks the wide range of engineering efforts that are required to launch a commercial product. As the design space grows, it becomes almost impractical to design at the device level. In CMOS technology, there are numerous tools from multiple vendors available commercially that allows designers to specify the desired functionality at a very high level without getting into the nitty gritty of the transistors. These tools are extremely trustworthy and effective. Such robust tools are not available yet for superconducting technology. Available tool chains for superconducting technology are not fully developed yet. In ~\cite{xu2017synthesis}, the authors propose a design flow for AQFP circuits, that includes logic synthesis, semi-automatic routing and HDL-based back-end veriﬁcation. A synthesis tool for RQL circuits is proposed in ~\cite{shauck2017reciprocal}. Muchuka et al in ~\cite{muchuka2016superconducting} propose a complete tool chain for superconducting electronics by evaluating and selecting the most suitable open-source tool for every stage in the design flow for small-scale, medium-scale, and large-scale circuits. A history of superconducting electronic design automation, roadmaps, basic design flows, requirements and challenges are described in ~\cite{fourie2018digital}, with focus on RSFQ, E-RSFQ and AQFP circuits. The IARPA SuperTools program was started in 2016 to develop a comprehensive set of tools that increase the scale, efficiency, and manufacturability of superconducting designs.

Ever since the idea of deploying Josephson junctions for digital logic popped, multiple circuit families have been proposed for superconducting technology. These include RSRQ ~\cite{likharev1991rsfq}, LV-RSFQ ~\cite{tanaka2013low}, ERSFQ ~\cite{kirichenko2011zero}, ADQP ~\cite{takeuchi2013adiabatic}, RQL ~\cite{herr2011ultra}, to name a few. A comparative analysis of these families are described in ~\cite{soloviev2017beyond}. A large number of circuits were demonstrated in RSFQ logic in the 1990s period, including digital signal processors ~\cite{gupta2007digital}, microprocessor components ~\cite{mukhanov1993rapid, mukhanov1993rsfq, filippov20118, dorojevets20138, yamanashi2007design}, mixed signal devices ~\cite{rylov1995superconducting, inamdar2009progress, kirichenko2003multi}, floating point units ~\cite{mukhanov1995implementation, kainuma2011design}.

A Josephson junction based processor was proposed as early as 1980 ~\cite{anacker1980josephson}. But due to slow progress of the technology and lack of engineering advances, the technology is under development even today. The Cryogenic Computing Complexity (C3) program was started by the Intelligence Advanced Research Projects Activity (IARPA) in 2014 to accelerate the growth of this technology  ~\cite{manheimer2015cryogenic}. The program proposes to demonstrate a superconducting computer prototype with a target frequency of 10 GHz, to meet the growing demands of exascale computing.

Superconducting technology is based on the principles of quantization of magnetic flux in a superconducting loop. It means that the magnetic flux ($\phi$) in a superconducting loop can only take integer multiple values of a single flux quanta (SFQ) ($\phi_0$) of magnetic field. Presence or absence of SFQ can thus be used to represent digital information. Superconducting circuit families are developed using Josephson junctions. A Josephson junction (JJ) consists of two superconductors coupled by a weak thin insulating barrier. Superconductivity is a physical phenomenon, in which certain metals exhibit no-resistance to the flow of electric current at very low temperatures (near absolute zero). The maximum super-current that can flow through the superconductor is called critical current. By controlling the current flowing through the Josephson junction, it can be switched back and forth between superconducting and resisitive states. This facilitates digital logic operations in superconducting logic. The critical current of a JJ is directed impacted by operating temperatures. The most commonly used Nb based JJs can operate in the thermal range of 4.2K ~\cite{han1987temperature}. Thus, systems built using superconducting technology require commercially available refrigeration. Since the introduction of superconducting technology, various circuit families have been proposed such as RSRQ ~\cite{likharev1991rsfq}, LV-RSFQ ~\cite{tanaka2013low}, ERSFQ ~\cite{kirichenko2011zero}, ADQP ~\cite{takeuchi2013adiabatic}, RQL ~\cite{herr2011ultra}, to name a few. A comparative analysis of these families are described in ~\cite{soloviev2017beyond}. 
}

\section{Superconducting Accelerator}

Superconducting circuits offer high energy efficiency. However, with limited device density and memory capacity, designing superconducting general purpose computers is incredibly challenging. Furthermore, lack of sophisticated design tools exacerbate the density problem as existing CMOS-based synthesis tools used for JJ design can not maximize device utilization. We believe that both of the problems are related to design and manufacturing economics rather than being fundamental challenges. However, until the technology reaches the maturity to manufacture and test billion Josephson Junctions per cm$^2$, which is typically required for general-purpose computing, we can leverage the technology to build accelerators. 

Applications with tiny working set size and high computational intensity are ideally suited for JJ-based accelerators. We study the SHA-256 application for building accelerators using JJ technology. We provide an overview of the application, the baseline CMOS implementation, and our JJ-based implementation.  We optimize the JJ design for performance (in Section~\ref{sec:tech_aware}) and reliability (in Section~\ref{sec:reliability}). We use the methodology and workflow described in Section~\ref{sec:methodology} for evaluations using the JJ technology.


\ignore{
Applications with tiny working set size and high computational intensity are ideally suited for JJ-based accelerators.  We study a SHA-256 accelerator, as it is commonly used for bitcoin mining. We choose to design a SHA accelerator due to its simple yet rich design that enables unique design trade-offs offered by JJ technology. This section provides a brief background of bitcoin mining and SHA-256 algorithm, followed by conventional accelerator for SHA-256, and our basic design based on JJ technology. Finally, we estimate the JJ complexity, performance (GH/s), and energy efficiency (GH/J).}
\vspace{-0.05 in}

\subsection{Background on Bitcoin-Mining}
\begin{figure}[!thb]
    \includegraphics[width=0.9\columnwidth]{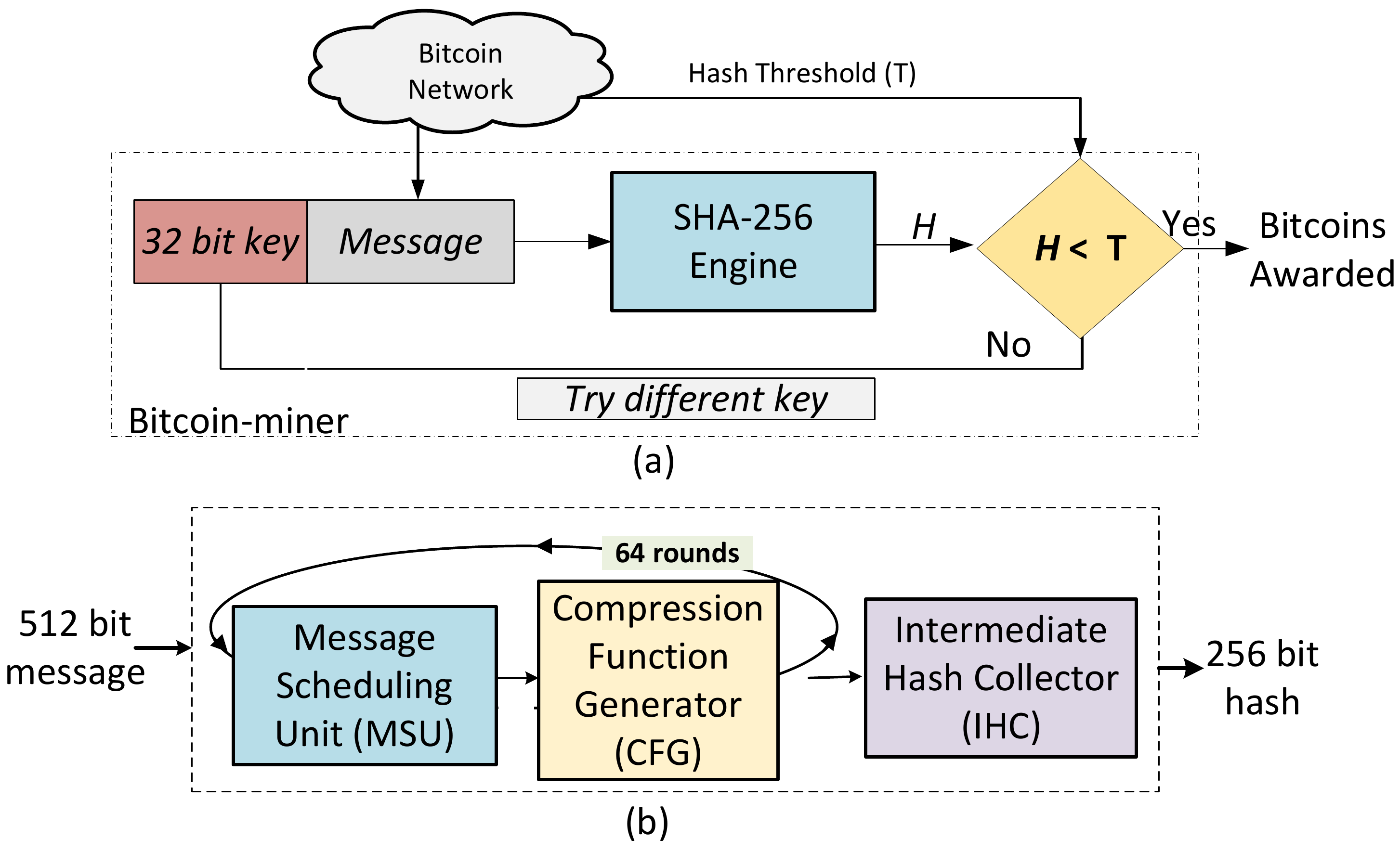}
     \vspace{-0.15 in}
    \caption{(a) An overview of Bitcoin Mining  (b) Overview of the SHA-256 Algorithm}
   \vspace{-0.15 in}
    \label{fig:sha_algo}
\end{figure}

A \texttt{blockchain} is a decentralized public ledger of transactions that maintains the validity of transactions by a distributed consensus mechanism~\cite{nakamoto}. In bitcoin, the process of authenticating transactions in this public ledger is called \texttt{mining}. It involves searching for a 32-bit key known as \textit{nonce} value such that when combined with the message which lists the transactions, the double \texttt{SHA-256} hash of the block (message + key) falls within a certain range. Whenever a miner finds a block i.e. 32-bit key (nonce) for an input message that leads to desired hash, the miner is rewarded with bitcoins. The overall process for bitcoin mining is captured in~\ref{fig:sha_algo}(a).
A bitcoin miner tries to maximize profit by trying multiple keys in parallel and as fast as possible as the probability of finding the key and getting rewarded is directly proportional to the total \texttt{hash rate}. However, repeated SHA-256 computation requires substantial power due to high computational intensity. The net profit depends both on the reward and operating costs (energy consumption~\cite{o2014bitcoin}). Hence, \texttt{energy-efficiency} (in GH/J) is the \texttt{figure-of-merit} that is optimized to increase profits. For this reason, bitcoin mining has evolved from CPUs to GPUs to FPGA and finally to ASIC-based implementations in the last decade ~\cite{magaki2016asic, taylor2017evolution}. 


\subsection{Background on SHA-256 Algorithm}


The SHA-256 computation of a message is carried out as shown in Figure~\ref{fig:sha_algo}(b). The message scheduler unit (MSU) takes an incoming message and splits it into 512-bit chunks. The MSU schedules a different 32-bit data to the compression function generator (CFG) every cycle, consuming 512-bit data over 64 rounds. The CFG uses this data and predefined constants to generate a 256-bit intermediate hash after every 64 iterations which is collected by the intermediate hash collector (IHC). When the entire message is processed, the values in the IHC registers is the final 256-bit hash.

 \begin{figure*}[!ht]
\centering
\vspace{-0.1 in}
    \includegraphics[width=1.9\columnwidth]{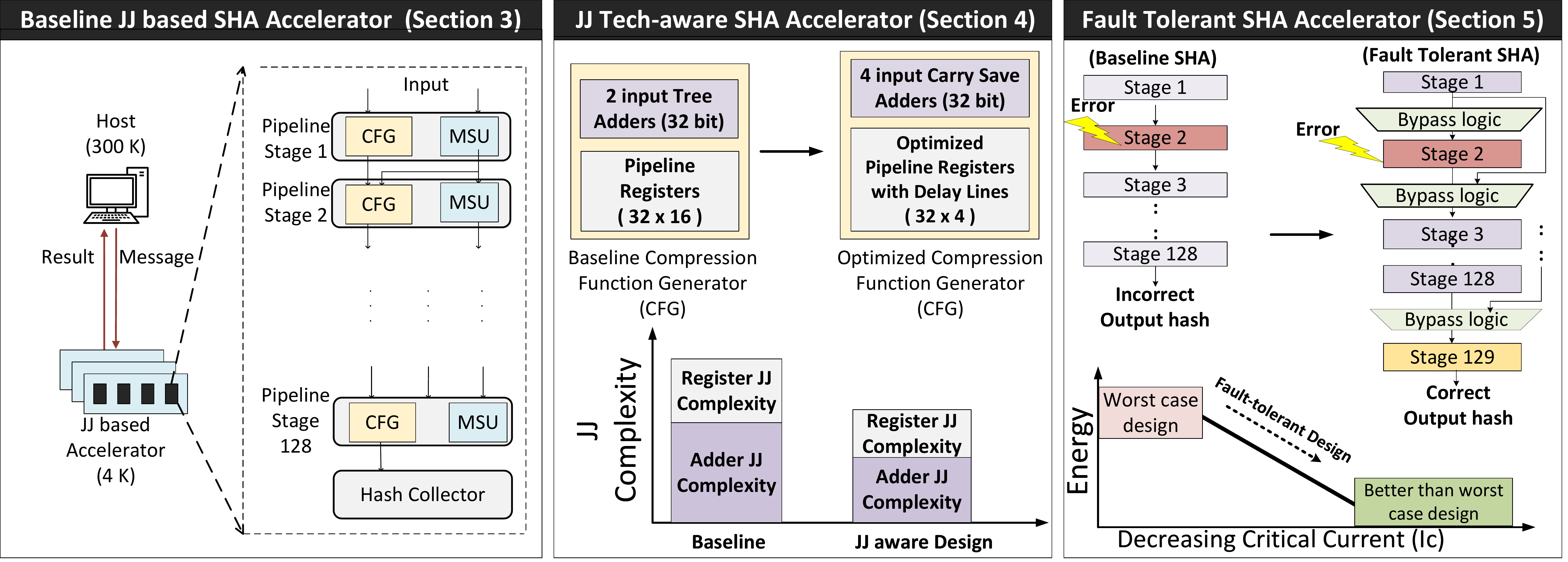}
    \caption{Design of a JJ-based Accelerator for SHA-256 based on GoldStrike 1 (a) Host at 300K communicates with accelerator at 4K  (b) Technology-aware design (c) Fault tolerant design}
    \label{fig:cold_acc}
    \vspace{-0.1in}
\end{figure*}

\subsection{Baseline CMOS Accelerator Design}
Bitcoin mining ASICs are available commercially from different vendors today. Furthermore, the state-of-the-art ASICs are fully custom designed at 16 nm or lower technology nodes and implement several design and algorithmic optimizations to increase the throughput (GH/s) and energy efficiency (GH/J). However, bitcoin mining is a competitive industry and the designs of state-of-the-art industrial accelerators are often kept proprietary.  In order to make a technology comparison for the same accelerator design, we use the publicly available Goldstrike1~\cite{Goldstrike} miner as the baseline for our studies (we compare our proposal with the publicly reported energy-efficiency for 16 nm AntMiner S9 in Section~\ref{sec:evaluation}). 

A hash engine contains  two instances of the SHA-256 computation blocks. SHA-256 algorithm uses 64 iterations, which can be pipelined. In Goldstrike1, these iterations are fully unrolled for both the rounds that eventually lead to a 128-stage pipeline.  Each pipeline stage comprises a compression function generation (CFG) logic and a message scheduling unit (MSU). The hash collector compares the output hash with the \texttt{target} to be achieved and if the criterion is met, it sends the result to the host.
\vspace{-0.01 in}


\subsection{Superconducting Accelerator Design}

We propose a superconducting blockchain accelerator that operates at 4K temperature and communicates with a host at room temperature. The architecture of our hash engine is shown in Figure~\ref{fig:cold_acc}(a).  The host receives the incoming messages from the network and offloads them to the accelerator. The accelerator computes hashes for different nonce values and it sends a message to the host when the network target is met. We port the CMOS Goldstrike1 design to superconducting logic without any optimization. 

SHA-256 algorithm requires computation using predefined constants. In our fully pipelined design, each pipeline stage requires a different fixed 32-bit constant for the computations, which are tied-off in the superconducting design to save on resources. The rotations and shifts in the SHA-256 computation involve fixed rotate/shift amounts. So the design does not implement any actual rotator or shifter logic but requires the signals to be routed appropriately. 


\subsection{Design Overview}

\begin{figure}[!b]
 \vspace{-0.15in}
\centering
    \includegraphics[width=0.8\columnwidth]{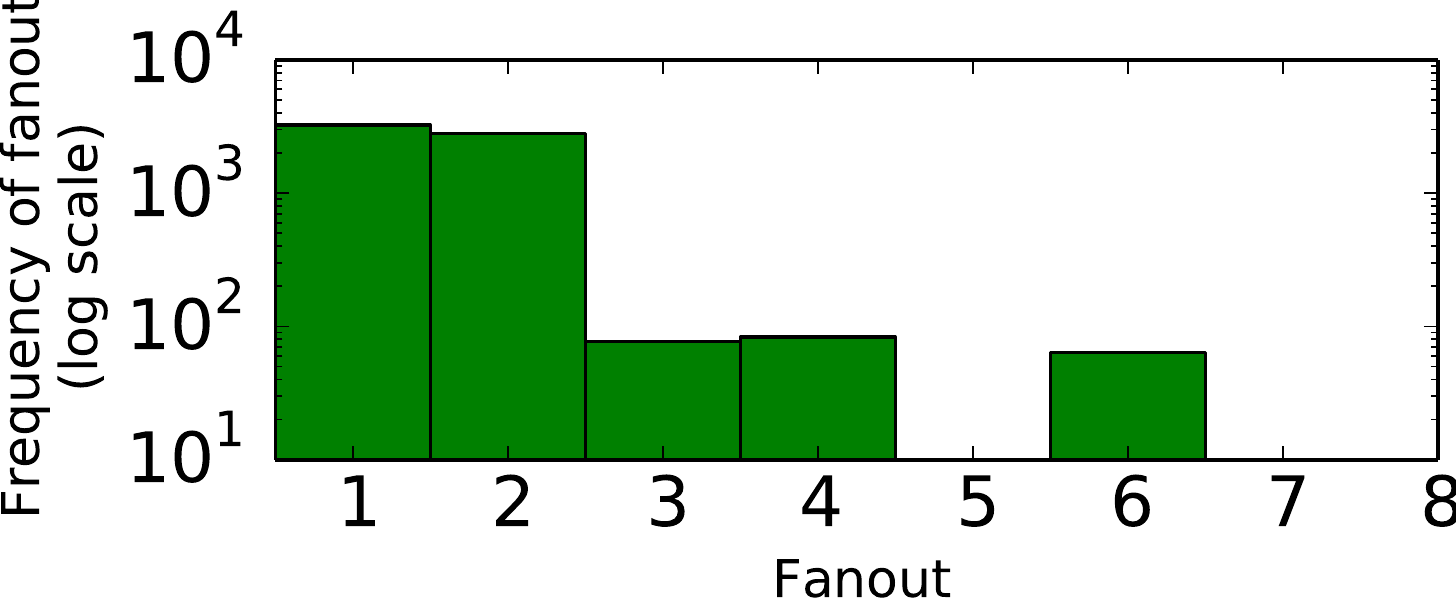}
    \vspace{-0.15in}
    \caption{Distribution of fanout in a single pipeline stage of baseline SHA accelerator}
 \label{fig:fanoutdis}   
   \vspace{-0.15in}
\end{figure}

Figure~\ref{fig:cold_acc}(a) shows an overview of our JJ-based implementation of GoldStrike1, which is designed by simply porting the CMOS-based implementation to JJ-based implementation.  Based on our methodology described in Section~\ref{sec:jj_modeling}, we compute the area (measured in JJ-complexity) for this design. The baseline design incurs significant JTL overheads (buffers that are required to facilitate fanout). 

For the analysis of JTL overheads, we study the distribution of fanout in our design. Figure~\ref{fig:fanoutdis} shows the distribution of fanout in a single stage of our hash engine. Thus, a gate drives on average about 1.5 gates, requiring 50\% additional JJs for fanout, incurring significant area overheads.

We perform a design space exploration to best meet the requirements of superconducting technology and present our results for a technology aware design of the superconducting SHA accelerator in Section~\ref{sec:tech_aware} (as shown in Figure~\ref{fig:cold_acc}(b)). Reliability is a key challenge in superconducting logic circuits and we present a case for a reliable, fault tolerant SHA accelerator in Section~\ref{sec:reliability} (as shown in Figure~\ref{fig:cold_acc}(c)).

\subsection{Performance and Energy-Efficiency}
In our design, 128 different values of nonce are processed in the pipeline and a hash is generated every cycle once the pipeline is full. The critical path in our design comprises of four adders in the CFG. 
We report the hashrate, power, and energy-efficiency in Table ~\ref{tab:perfbased} for the accelerator using the methodology described in Section~\ref{sec:methodology} for two design points, with ripple carry adders (RCAs) and Kogge-Stone adders (KSAs). An RCA is 3x more energy-efficient than a KSA but a KSA has 30\% lower latency. This enables us to compare two design points, one that is optimized for energy-efficiency and another that is optimized for performance. For the high performance design, KSAs are used economically since they are expensive in terms of resources. They are used only to optimize the speed-path and the non-critical path adders are still designed to be RCAs. 
Table~\ref{table:perfbased} also compares the performance and energy-efficiency of the GoldStrike 1 accelerator designed with superconducting logic using the baseline CMOS-based architecture for the two different design points.

\begin{table}[htp]
\centering
\label{table:perfbased}
\begin{small}
\vspace{-0.15 in}
\caption{Performance and Energy Evaluations for SHA accelerator implemented in CMOS and JJ}
\renewcommand{\arraystretch}{1.2}
\label{tab:perfbased}
\begin{tabular}{ |c|c|c|c| } 
\hline
Parameter                               & GoldStrike 1 &              JJ-Design          &       JJ-Design          \\
                                      &  CMOS  & only RCA  & with KSA \\ \hline \hline

Technology                              & 16 nm                            &    248 nm         &   248 nm              \\
\hline

JJ Complexity (million)                          &   N/A        &   3.38 &   5.54 \\
\hline

%
%
Hashrate (GH/s)                         & 1.05                                    &    0.661                  &       0.951                   \\
\hline
%
%
%
%
Total Power (milli-Watt)                         &  250                                     &    15.65      &       36.23           \\
\hline
\textbf{Energy-Efficiency}                 &  \textbf{4.0}                                  &    \textbf{42.26}                  &       \textbf{26.24}                   \\

\textbf{(GH/J)}                &           (1x)                      &       (10.6x)               &      (6.56x)                    \\
\hline

\end{tabular}
\label{table:perfbased}
\end{small}
\end{table}

The JJ-based design that is implemented with only Ripple-Carry Adders is 10x more energy efficient than the CMOS implementation, however it has 37\% lower performance. Using Kogge-Stone adders reduce the energy-efficiency to 6.5x while bridging the performance difference to within 10\%. We observe that our design energy-efficiency reduces by almost one-third for design optimized with KSAs, indicating that optimizing only for high-speed can be detrimental to the overall energy-efficiency. However, both designs show that simply porting the accelerator from CMOS to superconducting logic can provide significant energy-efficiency improvement. 

The contribution towards JJ-complexity for our hash engine comes from adders, registers and other logic. Table~\ref{tab:jjcomplex} shows the contribution towards JJ-complexity from each of these three sources. We  observe that about 50\% of the contribution towards JJ-complexity is from adders for an RCA-based design and this increases to 67.7\% for KSAs. Optimizing the accelerator design to suit the specific constraints of the JJ technology can further improve energy efficiency. We discus technology-aware optimizations in Section~\ref{sec:tech_aware}.

\begin{table}[!hb]
\centering
\begin{small}
\caption{Breakdown of JJ-complexity}
\setlength{\tabcolsep}{0.05cm} 
\renewcommand{\arraystretch}{1.2}
\label{tab:contribution}
\begin{tabular}{ |c|c|c|c|c| } 
\hline
Design & Adders & Registers & Other Logic & Total  \\
       & (million) & (million) & (million) & (million)\\ 
\hline
With RCAs & 1.69 (50.1\%) & 1.51 (44.8\%) & 0.17 (5.1\%) & 3.38\\
\hline 
With KSAs & 3.75 (67.7\%) & 1.51 (27.3\%) & 0.28 (5.0\%)  & 5.54\\
\hline
\end{tabular}
\label{tab:jjcomplex}
\end{small}
\end{table}


\section{Technology-aware Design}
\label{sec:tech_aware}
In this section, we discuss the impact of JJ technology on design and architectural decisions. To illustrate the contrast between CMOS and JJ designs, we focus on two critical components of the SHA engine: adders and registers. We also discuss a way to optimize communication for the accelerator.

\subsection{Tradeoffs in JJ Adder Circuits}
The proposed SHA engine uses 1200 adders, which accounts for more than 50\% of JJ-complexity. Furthermore, the clock frequency of the SHA engine is dictated by the critical path that consists of four additions in CFG unit as shown in Figure~\ref{fig:cold_acc}(b). Adders dominate the on-chip resources and overall latency. Thus, optimizing the adders to improve critical path and overall energy efficiency is essential.

\begin{figure}[t]
\centering
    \includegraphics[width=0.9\columnwidth]{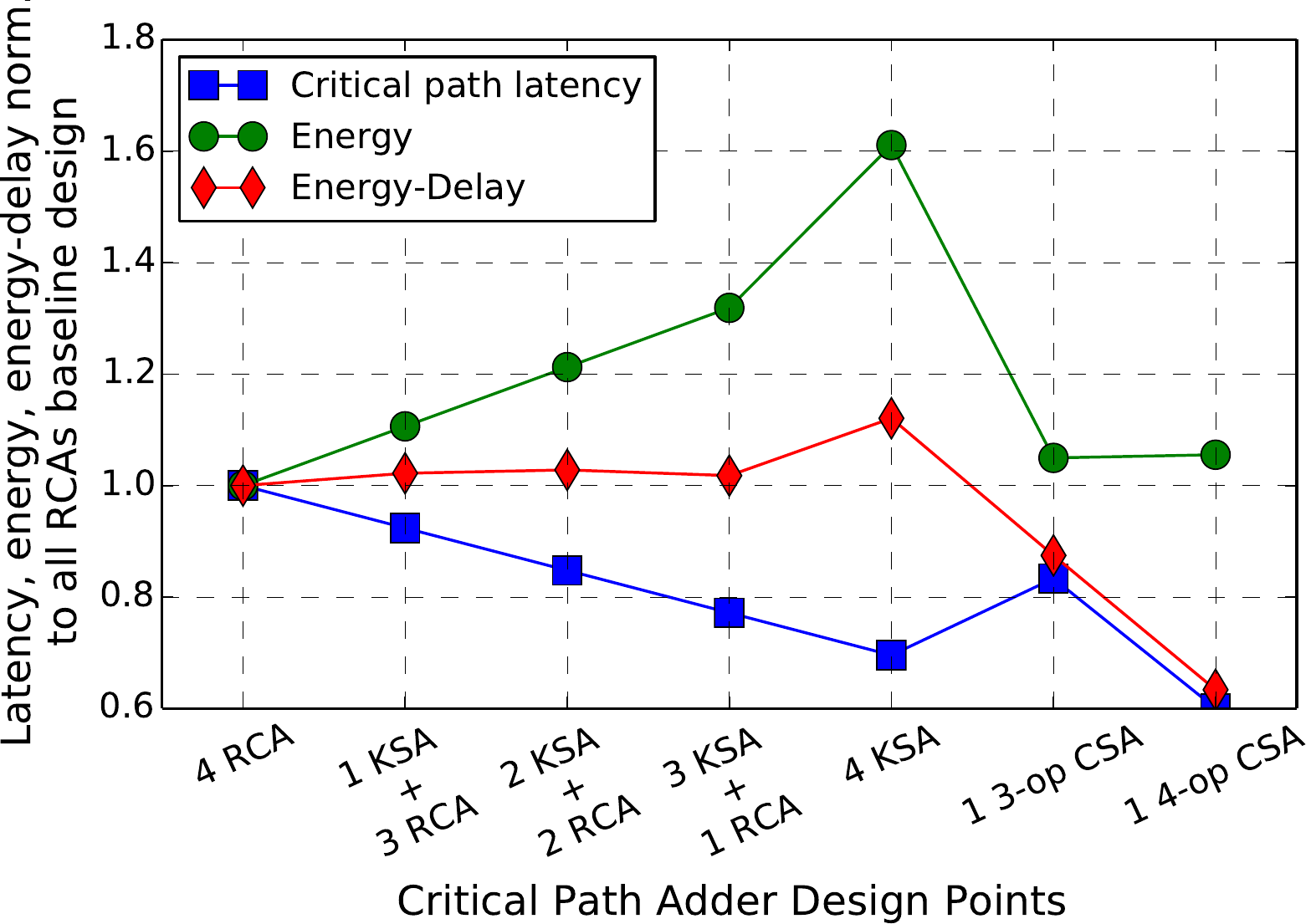}
    \caption{Latency, energy, and energy-delay product for different critical path adders designs normalized to Ripple Carry Adder (RCA) parameters}    
    \label{fig:energy_delay}
    \vspace{-0.10in}
\end{figure}

Typically CMOS adder designs improve latency at the expense of more transistors or complex connectivity. Although, the complexity of the adder design increases, the delay and energy efficiency also improves. For example, a complex {\em Kogge Stone Adder (KSA)} is faster and more energy efficient as compared to  simple {\em Ripple Carry Adders (RCA)}. However, JJ based adders do not follow the same trends. For instance, tree based adders rely on complex communication patterns to improve the critical path from $O(N)$ to $O(log_2{N})$. However, to enable tree based adders, we need greater fan-out and complex wiring, both of which have low overhead in CMOS. However, the limited fanout of RQL gates require JTLs leading to high JJ complexity. For example, JJ based KSA improves performance, but it worsens the energy-efficiency~\cite{dorojevets2015towards,dorojevets20138}. While designing the SHA engine, a combination of adders can be selected such that our design meets the baseline CMOS performance and maximizes overall energy-efficiency. To satisfy these criteria, we choose different design combinations of KSA and RCA as shown in Figure~\ref{fig:energy_delay}.


\subsection{ Fanout-aware Adder Design}

Table~\ref{table:perfbased} show that merely replacing RCAs with KSAs improve the critical path but degrades the energy- efficiency. Furthermore, even after replacing all four critical path RCAs with KSAs, JJ based accelerator fails to meet the baseline delay. Our goal is to meet the critical path requirement without deteriorating the energy efficiency. Thus we try to optimize our design such that JTL overheads are reduced, and simplicity of RCA is maintained. We observe that majority of the additions in the CFG of Figure~\ref{fig:cold_acc}(b) are back to back additions and most intermediate addition results are not used elsewhere. Thus, it is possible to replace some of these adders by a sequence of carry save adders (CSA). An n-operands CSA computes the composite addition much faster as compared to ripple carry adders. If $\delta_{FA}$ is the delay of a 1-bit full adder (FA), the latency of an N number addition with CSA that can add $N$ $k$-bit numbers is given by $Latency_{CSA} = (K+N-1)\delta_{FA}$.

\ignore{
\begin{figure}[!htb]
\centering
    \includegraphics[width=0.75\columnwidth]{Figure/Adder.pdf}
    \caption{ Four operand (s = a+b+c+d) modulo-32 carry save adder schematic}
    
    \label{fig:faa}
\end{figure}
}

CSA has lower fanout and does not requires routing between distant gates, making it more layout friendly. When two back to back adders on the critical path of the CFG and the MSU are replaced by a 3-op CSA, the design has 1.2x the performance and is 1.25x more energy-efficient than our baseline design, even after accounting for 20\% skew. Hardware optimizations have been proposed in the past to move the addition of variables $W_i$ and $K_i$ from MSU to CFG ~\cite{ting2002fpga}. We propose a similar optimization where this value is pre-computed in the $(i-1)^{th}$ stage of the pipeline and consumed in the $i^{th}$ stage. This allows us to use 4-op CSA in both CFG and MSU blocks and fuse 3 additions, thereby reducing the overall critical path. This design offers 1.67x the performance improvement for RCA baseline design and is 1.44x more energy-efficient. Table~\ref{tab:rqlhashl} lists the performance and energy-efficiency of the superconducting hash engine for the different adder optimized designs against the baseline design using all RCAs.  A similar optimization uses multiple such CSA in parallel for a high-speed SHA-256 ASIC design besides carry-lookahead adders~\cite{dadda2004design}. However, we use these CSAs in conjunction with energy-efficient RCAs to have a more economical design in terms of JJ-complexity.

\subsection{Reducing Registers Using Delay-Line}

In the baseline pipeline design, in each stage MSU uses 16 registers with 32-bit width, and  CFG uses 8 registers. This results in about 35\% JJ complexity for an optimized adder circuit. The registers hold the input values and the intermediate results. The contents of the registers are consumed by adders and other logic to produce an output in every stage, and subsequent stages consume the produced output of the stage. A baseline design replicates all the 24 registers at every stage requiring large JJ complexity.  Furthermore, all the register values from $i^{th} $ stage to $(i+1)^{th}$ stage are expected to flush every clock cycle. We would need a wide bus to flush the contents of registers every cycle. wide bus and a small set of registers are trivial in CMOS. However, in JJ technology, JTL cost of wide buses and registers lead to high costs.

\begin{figure}[!t]
\centering
    \includegraphics[width=0.9\columnwidth]{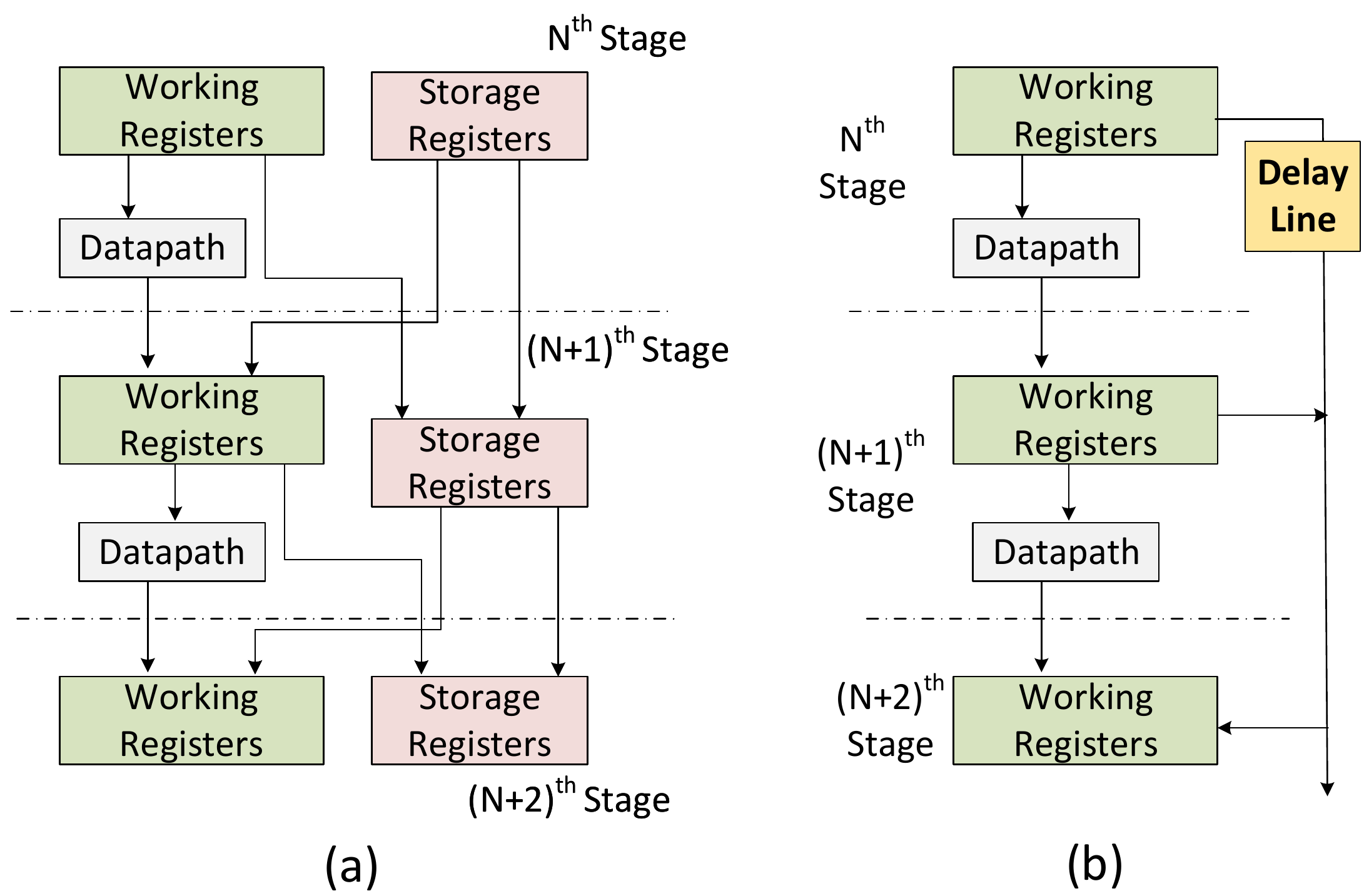}
    \vspace{-0.15in}
    \caption{(a) Basic design needs 16 registers (storage + working registers) in MSU per stage (b) Optimized design with delay-line reduces it to 4 working registers}
    \label{fig:delay_line}
\vspace{-0.1in}
\end{figure}

In traditional non-pipelined SHA-256 design a global register holds all the intermediate values. Whereas, in the quasi-pipelined SHA designs (SHA engine with 4 stage pipeline) each pipeline stage uses a local register~\cite{dadda2004asic}. The local register file enables a higher clock rate. Whereas, shared registers improve the critical path significantly, especially for heavily pipelined designs requiring data values every cycle. So, to supply register values each uses a local set of registers leading to high JJ complexity (registers account for 35\% JJs).

The baseline design has fixed control path, and identical operations are performed in every stage of the pipeline. Each pipeline stage produces an output that is consumed in the next set of stages. For example in MSU, only  four registers are consumed by 4 input 32-bit adder in each stage of the pipeline, to produce one 32-bit output. After that, all the registers are simply copied to the next stage, such that $N^{th}$ register of the current stage is copied to $(N+1)^{th}$ register of the subsequent stage as shown in the Figure~\ref{fig:delay_line}(a). Thus only one register value is produced, four values are consumed and rest of the register values are copied as is to the next stage. We can leverage this deterministic production and consumption of the values to eliminate the large fraction of registers.   

\begin{figure}[!b]
 \centering
 \vspace{-0.1in}
    \includegraphics[width=0.90\columnwidth]{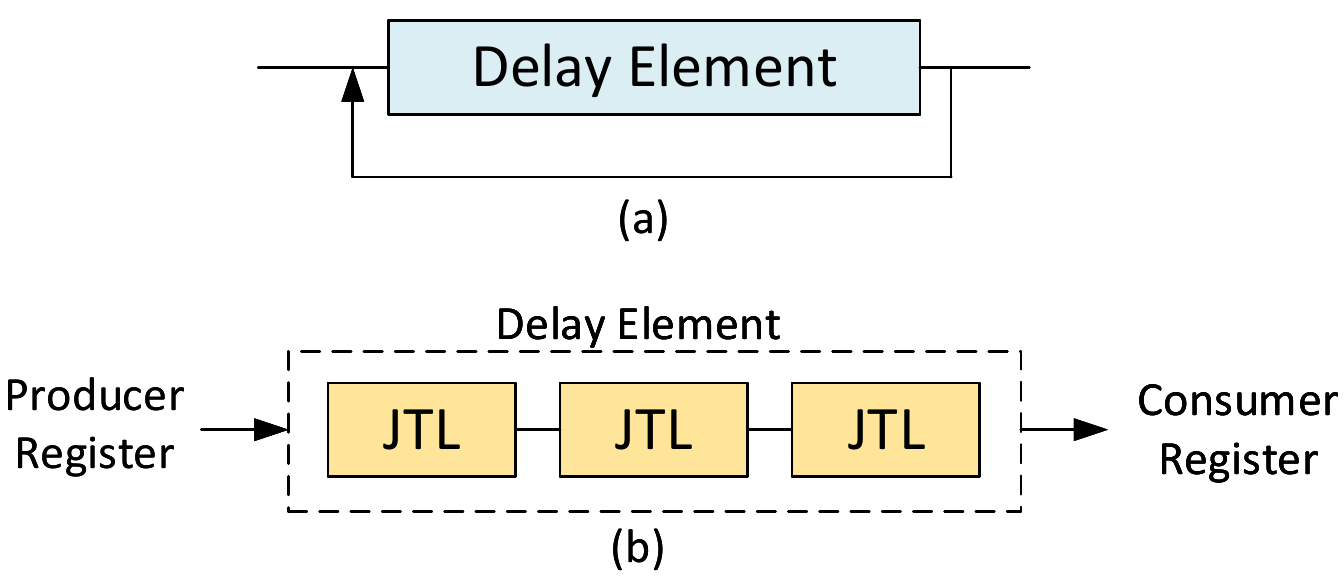}
    \caption{(a) Delay Line Memory (b) JTL based delay line can delay and forward data from one stage to other}
    \vspace{-0.1 in}
    \label{fig:dlm}

\end{figure} 
An alternative to communicating between stages using registers is to connect producer and consumer via a {\em Delay-Line}. Delay line memory is a form of memory used in earliest computers during 1960s~\cite{eckert1953memory,auerbach1949mercury}. Unlike modern day random access memories, a delay line memory is based on sequential access and requires to be refreshed from time to time. Such memories rely on transmitting information through a circuitry that adds delay and re-routing the end of the delay path to the input end such that information can be transmitted continuously through the closed loop as shown in Figure~\ref{fig:dlm}(a). We propose to delay lines to route data from producer register to consumer register in a synchronized manner by using the precise number of delay elements to match the desired delay.

In RQL, a delay line can be built using a series of JTLs that repeat signals for every clock activation. On a JTL delay line, input data is propagated from one JJ to next JJ  every clock phase. This provides an efficient way to leverage the producer and consumer patterns in a hash engine to reduce JJ complexity. Delay lines keep the data in flight and deliver to the consumer at the precise clock cycle. Since the delay line can simply load a new value every clock cycle, it can be integrated seamlessly with the proposed pipeline design.

A delay line can facilitate the delivery of intermediate results from a producer stage to a consumer stage. The cost of delay line memory is 4 JJs per clock cycle per bit whereas register storage requires 12 JJs per bit. Although the crossover-point for the flop based register file is 3 clock cycles, the delay line memory enables point to point connection between the producer and consumer that eliminates the need for 16 registers for every stage. We use four staging registers along with the delay line design to tolerate clock skew. The delay lines reduce the per stage JJ cost by almost 20\%. Furthermore, it simplifies the bus design.

\subsection{Performance and Energy-Efficiency}

Table~\ref{tab:rqlhashl} shows the performance and energy-efficiency of our baseline JJ-based accelerator (with RCA/KSA) and with technology aware optimization of 4-operands CSA (four-input) and use of delay-lines to reduce register cost. The 4-operands CSA optimization improves the energy-efficiency from 6.39x for KSA to 10.0x, while also improving the performance by 15\% (bringing it in line with the performance of the CMOS-based implementation). The delay-line optimization reduces register file costs and improves energy efficiency from 10.0x to 12.4x, while still having similar performance.

\begin{table}[t]
\centering
\begin{small}

\caption{Performance and Energy after Optimization}
\vspace{-0.05 in}
\label{tab:rqlhashl}

\setlength{\tabcolsep}{0.05cm} 
\renewcommand{\arraystretch}{1.2}
\begin{tabular}{ |c|c|c|c|c| } 
\hline
Parameter                               & RCA        & KSA    &  4-CSA & 4-CSA + Reg     \\
                                        & Adder   &   Adder              & Adder      & optimization \\ 
                                        \hline 
                                        \hline
JJ Complexity (million)                 & 3.38              &  5.45       & 3.57       & 2.89 \\
\hline
Hashrate (GHz)                          &  0.661        & 0.951         & 1.101 & 1.101       \\
\hline
Total Power (mW)                         &  15.64         &  36.22       & 27.5 & 22.26    \\
\hline
Energy Efficiency (GH/J)                        &  42.27  &  26.24   & 40.05 & 49.47 \\
\hline
\textbf{Efficiency wrt CMOS-16nm}                        &  \textbf{10.56x}     &  \textbf{6.39x}     & \textbf{10.0x} & \textbf{12.37x}   \\
\hline
\textbf{Hashrate wrt CMOS-16nm}                        &  \textbf{0.63x}     &  \textbf{0.90x}     & \textbf{1.05x} & \textbf{1.05x}   \\

\hline

\end{tabular}
\end{small}

\vspace{-0.15 in}
\end{table}

\begin{figure*}[!ht]
\centering
    
    \includegraphics[width=2\columnwidth]{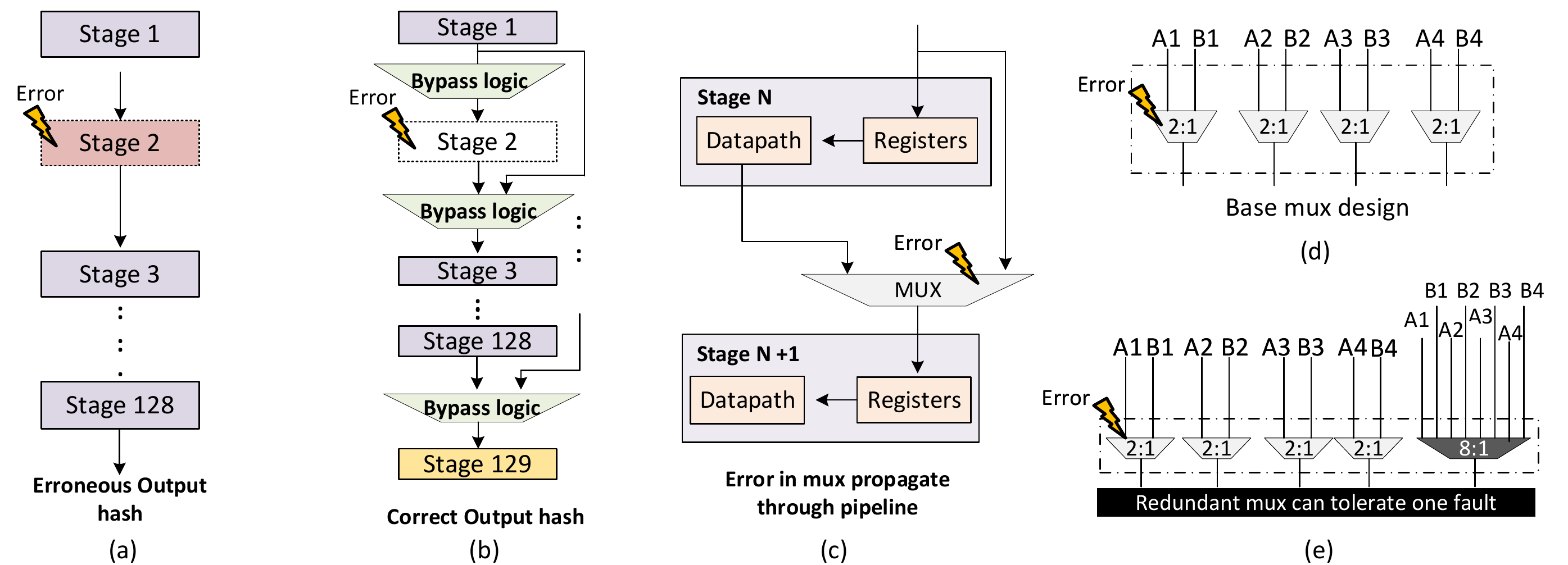}
\vspace{-0.05in}
    \caption{(a) Baseline design with no redundancy (b) Reliable design to mitigate correlated faults using sparing stage and bypass logic (c) Logic for bypass is vulnerable to faults (d)  Mux design for the selection logic for sparing technique (e) Reliable design using enhanced bypass logic with redundant mux that can tolerate at the most two faults.}

    \label{fig:realibility}
   \vspace{-0.1in}
\end{figure*}

\section{Fault Tolerant \& BTWC Design}
\label{sec:reliability}
In this section, we discuss fault models for JJ technology, and present a  design that can use architecture-level solutions to protect against these faults. We also discuss how the proposed fault-tolerant design can be leveraged to improve energy efficiency by operating the circuit  at a Better-Than-Worst-Case (BTWC) design point. 

\subsection{JJ Logic Fault Sources \& Models}

There are three primary sources of faults in superconducting logic: Fabrication defect,  device level variations, and non-ideal operating environment. Fabrication defects result from the material and masking defects introduced during fabrication. These defects can manifest as permanent stuck-at-faults, similar to birth-time defects in CMOS, and can be mitigated by design time testing. Device parameter variation can cause degradation in noise margins. For example, variation in critical current ($I_{c}$) can cause degradation in noise margin resulting in timing errors and the design must operate at a point where it is robust against such variations. 


In JJ technology, flux trapping causes a unique source of faults which we term {\em the operating environment fault}. These faults are challenging to protect against due to correlated nature of the errors. Furthermore, the faults are neither a permanent fault nor a transient fault, and it manifests not only because of the device but also due to non-ideal operating conditions. Flux-trapping results from trapping of stray magnetic field in the JJ circuits due to non-uniform cooling and can result in non-functional circuits or reduced noise margin for parts of the chip. Fortunately, steady progress and innovations in fabrication and device technology limits the problem of flux trapping considerably~\cite{herr2015reproducible}. The reported flux-trapping solutions are demonstrated on 50K JJ circuits. However, the techniques are costly, hard to scale to large systems, and do not completely eliminate the problem of flux-trapping. For example, some of the demonstrations use active magnetic field cancellation or extremely low temperature (<1K) at which flux vortex freezes, both of the additional requirements are expensive, especially for large scale systems. 



\subsection{Impact of Faults on SHA-256 Hardware}

To understand the impact of faults on the output of the SHA engine, we use fault injection to quantify the Architectural Vulnerability Factor (AVF). For the baseline design, injection of faults shows  98.89\% AVF.  The high AVF of SHA engine results from the entropy maximization property of the algorithm where a single bit operational error can corrupt the output. Protecting a SHA engine is a traditionally non-trivial problem due to its cryptographic properties and tight area and energy constraints. Techniques based on replication or parity detection circuits are either too complex and expensive or provide partial protection against faults.

\subsection{Application Level Resilience}

Transient faults do not have any meaningful impact on the mining process and hash-rate as transient errors can corrupt only one of the key combinations. The probability of a miner missing out on a reward due to a transient fault is extremely small. For instance, the probability of finding a block is relatively low ($\frac{1}{2^{32}}$), and  if the probability of transient fault is $P$, then collision of those two events is significantly lower ($\frac{P}{2^{32}}$). Recent proposals take advantage of this property to enable approximate bitcoin mining~\cite{vilim2016approximate}.

On the other hand, permanent faults would result in non-functional SHA engine thus reducing the yield significantly. Furthermore, if not detected before deploying, the miner would simply consume power without doing any work. This problem is significantly worse for the flux trapping faults as fault patterns change every warm up cycle which forces us to test SHA engines after every cool-down. In CMOS, non-functional chips can be isolated by post fabrication tests. Whereas, in JJ circuits, faults can happen not only because of fabrication defects but also due to operating conditions. 

\subsection{Fault Tolerant Design}

The correlated nature of faults due to large granularity impact of flux traps limit the ability to use standard low-cost protection techniques to protect against single-bit faults that happen in conventional technologies. Our goal is to leverage the regular structure of the accelerator to improve the reliability of the JJ based SHA-256 engine without significant complexity.

For the pipelined SHA-256 accelerator, all the stages in the pipeline are functionally identical. Furthermore, all the stages have deterministic control and data-path. This can be leveraged to enable low-cost fault tolerance. We propose to add an extra pipeline stage and build a bypass logic between consecutive pipeline stages such that if a fault is detected for a pipeline stage that stage can be bypassed as shown in Figure~\ref{fig:realibility}(b). The bypass logic and spare pipeline stage can be used to detect the faulty pipeline stage.  A faulty stage can be detected by bypassing the stages one by one  with a standard input and output pair until a right hash is produced. While we describe the solution with one spare pipeline stage, the same bypass network can be used to mitigate N fault stages in the accelerator by using N spare pipeline stages.

\begin{figure}[!tb]
\centering

    \includegraphics[width=1\columnwidth]{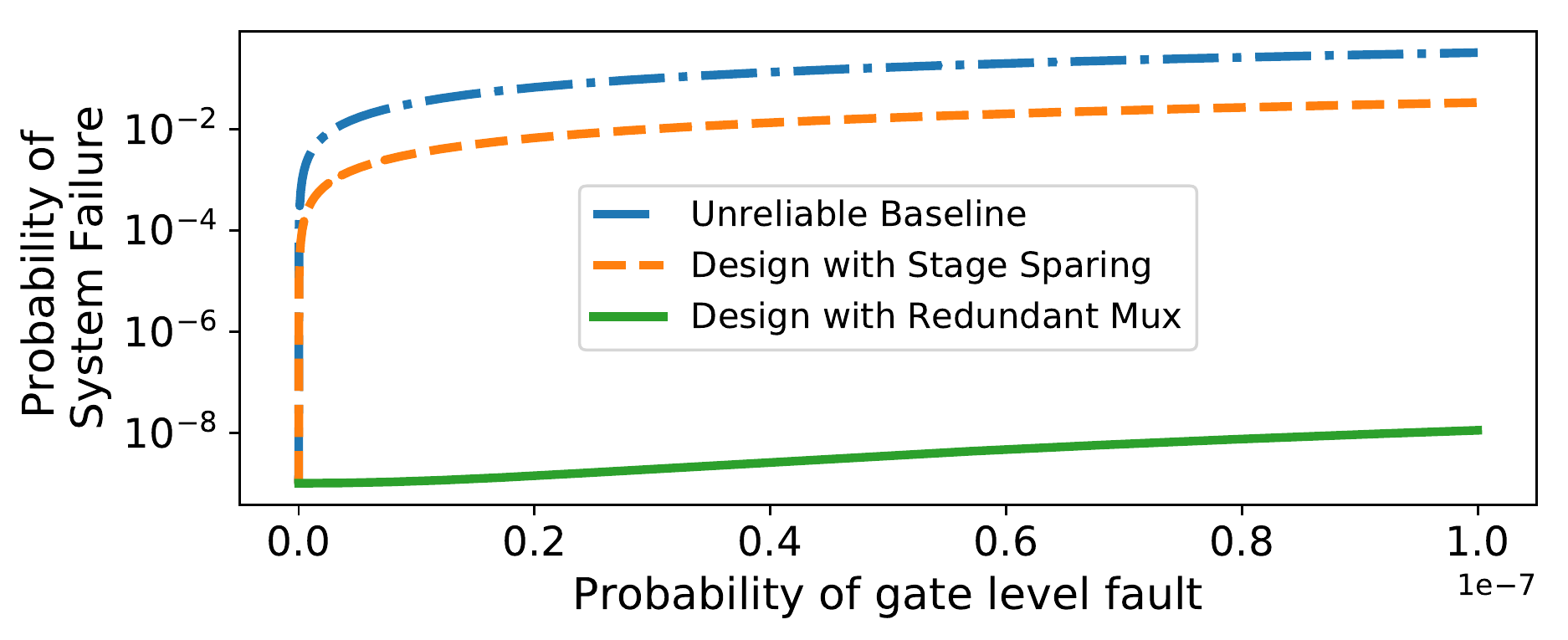}
    \vspace{-0.1in} 
    \caption{Probability of System failure with respect to probability of gate level fault due flux trapping for (a) baseline with no redundant structures (b) design with a spare stage and bypass selection logic (c) spare stage and redundant muxes}
    \vspace{-0.1in}
    \label{fig:psf}
 
\end{figure}

The bypass logic is placed between all 128 stages and it consists of four 32-bit 2:1 multiplexers as shown in Figure~\ref{fig:realibility}(c). The multiplexers can bypass the faulty stage and re-route the signals to subsequent working stages. The multiplexers are essential for routing signals from one stage to another even in the absence of faults as they are placed between two stages as shown in Figure~\ref{fig:realibility}(b). Fault on any of the multiplexers results in a non-functional SHA engine. However, multiplexers cover only a small fraction of total area and the likelihood of a fault occurring on any of the multiplexers is an order of magnitude less compared to other functional units. Thus, this design enables partial fault-tolerance as it can function correctly as long as faults do not occur on any of the multiplexer blocks.
To evaluate the effectiveness of the design, we perform  binomial trials assuming identical and independently distributed (iid) errors. 
In the baseline, even a single fault can lead to system failure whereas, sparing design build some fault-tolerance. To improve the reliability even further, we propose a spare stage design that uses a redundant multiplexers for bypass circuitry. As shown in Figure~\ref{fig:realibility}(e), the redundant multiplexer can tolerate one fault on any of the four multiplexers by using an extra 8:1 mux. The design with redundant mux can tolerate one fault anywhere. Figure~\ref{fig:psf} shows the probability of system failure for the baseline, stage-sparing, and stage-sparing with redundant muxes. The design with sparing and redundant mux is ~5-6 orders of magnitude more reliable as compared to the baseline.

\subsection{Using Fault-Tolerant Design for BTWC}

The energy-efficiency and performance of the superconducting circuit is determined by the critical current (Ic).  We can reduce the energy consumption by reducing Ic; however, this can cause certain devices to fail. Therefore, the critical current is set conservatively such that none of the devices fail. Recent studies suggest that the $I_{c}$ distribution for future technology nodes may have a large spread between devices, leading to as much as 5x difference between the average $I_{c}$ and worst-case $I_{c}$~\cite{holmes2017non}. The heavy tail on $I_{c}$ distribution may force designers to pick $I_{c}$ conservatively. However, we can leverage the proposed fault-tolerant design to tune the optimal $I_{c}$ by using a better-than-worst-case (BTWC) design philosophy. The proposed reliable SHA-engine design can be used to tune the $I_{c}$ as it can protect against a large granularity failure using a spare pipeline stage. To perform the run-time tuning, a $I_{c}$ value is lowered until a failure is observed. With the fault tolerant design, a weak pipeline can be detected and bypassed. If a fault can not be isolated, in that case the $I_c$ is increased. The tuning enables optimal $I_{c}$ by isolating a weak/faulty pipeline stage, and mitigation of the fault. This can reduce the $I_c$ from 38 $\mu A$ to 10 $\mu A$ (based on conservative scaling model of $I_{c}$ of Herr et al.~\cite{herr2011ultra}).

\subsection{Evaluations: Tying it All Together}
\label{sec:evaluation}
\begin{figure}[!htb]
    \includegraphics[width=1.0\columnwidth]{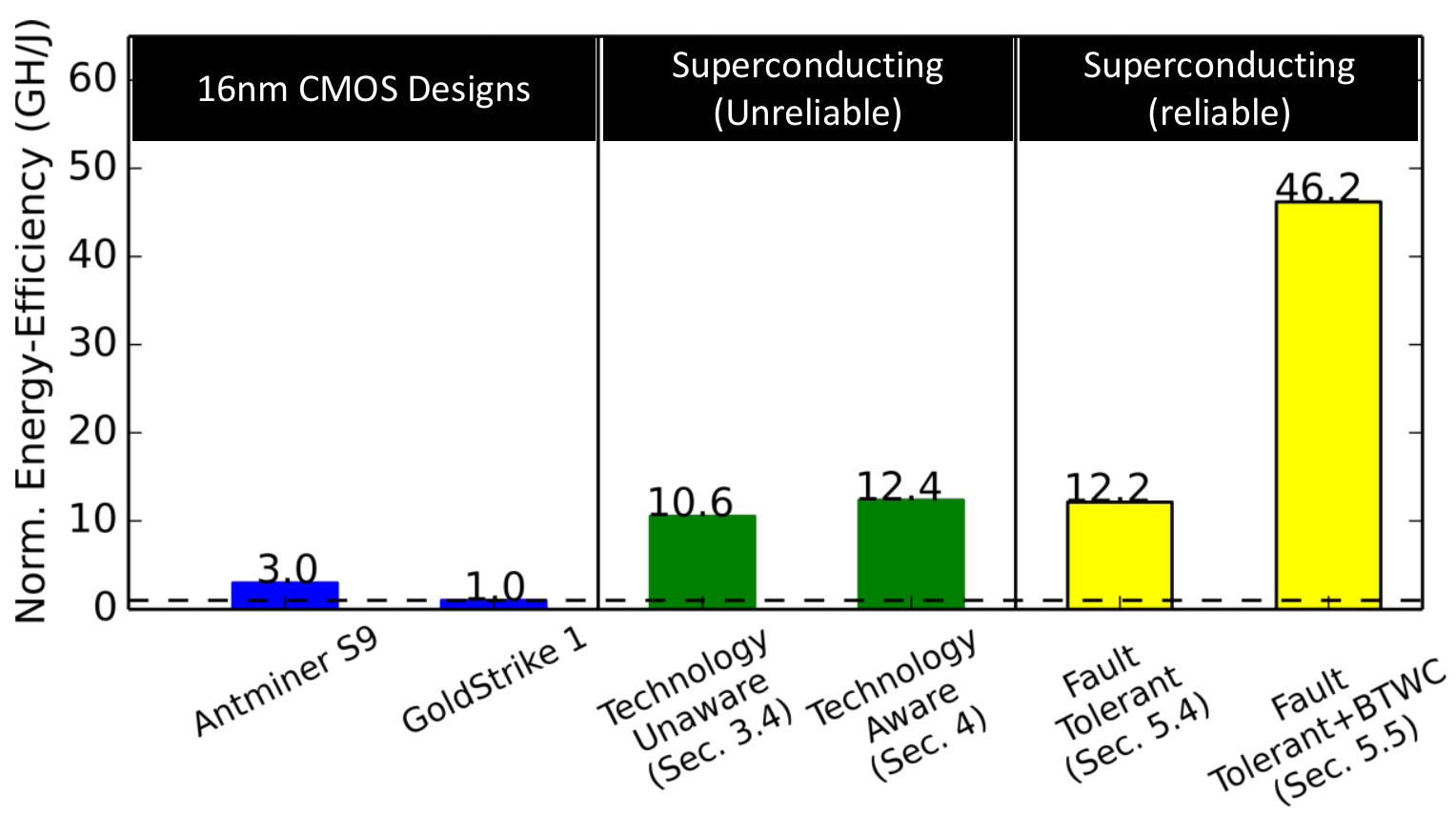}
    \caption{Energy-efficiency of CMOS and JJ-based implementations.  Our final design has 46x improvement over CMOS-based implementation. Note: All JJ-based evaluations include a cooling overhead of 300x. }
    \label{fig:eval}
    \vspace{-0.15in}
\end{figure}

Figure~\ref{fig:eval} shows the energy-efficiency of JJ-based designs, all normalized to the baseline CMOS implementation of Goldstrike 1. For reference, we also show the published numbers for commercial ASIC, Antminer-S9, which provides only a 3x improvement in energy efficiency. A basic design that simply translates the CMOS implementation of Goldstrike to JJ technology provides  10x improvement. Redesigning it for technology-specific constraints (fanout, efficient communication) improves the energy efficiency to 12.4x. To enable fault-tolerance,  a proposed fault-tolerant design with one spare pipeline stage has overall energy efficiency of 12.2x. The additional JJs required for bypass logic lowers the efficiency compared to unreliable design. However, the fault-tolerant design enables lowering of the critical current ($I_{c}$) from 38 $\mu A$ to 10 $\mu A$, increasing the energy efficiency to 46x (while having 1.2x the performance of the CMOS implementation).

\begin{figure*}[htb]
\centering
    \includegraphics[width=5 in]{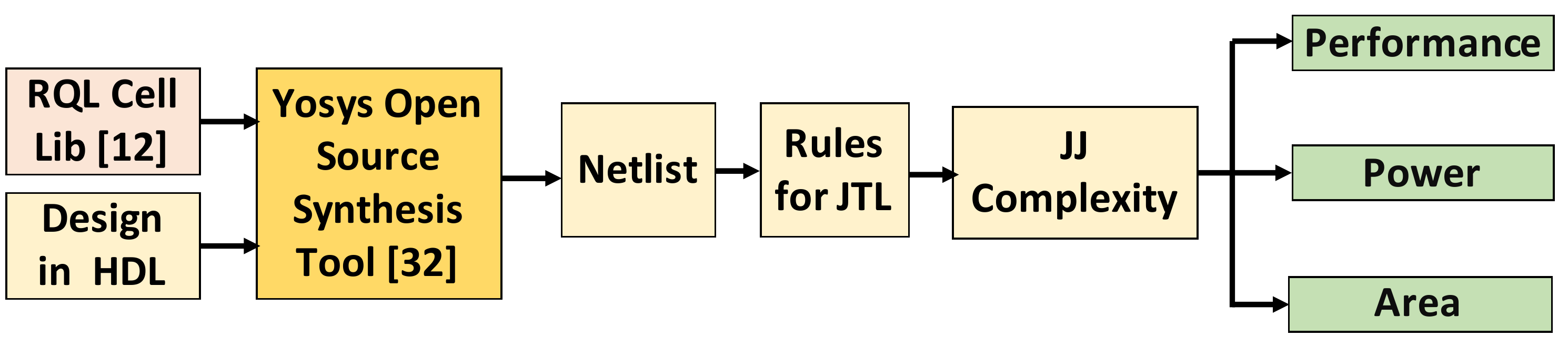}
    \vspace{-0.1 in}
    \caption{Workflow for evaluating area, performance, and power of superconducting accelerators.}
        \vspace{-0.1 in}
    \label{fig:workflow}
\end{figure*}

\section{Evaluation Workflow}
\label{sec:methodology}

To the best of our knowledge, this is one of the first paper to explore superconducting accelerators and evaluate the performance and power using application-level metrics.  As this is an emerging technology, there is no publicly available methodology or workflow for evaluating performance, power, and area of systems built using superconducting technology.  Furthermore, standard cells and design rules in superconducting logic families are fundamentally different from CMOS technology. For example, logic cells in JJ technology have limited driving capacity, and to drive more than one cell, a buffer like cell known as Josephson Junction Transmission Line (JTLs) must be placed between two cells. This limits the direct usage of standard CMOS tools to perform a design space exploration for superconducting technology. To overcome this problem, we use open-source back-end design tools to incorporate design constraints specific to superconducting logic. In addition to the modified design tool, we use analytical models to calculate performance, power, and area for superconducting logic. Figure~\ref{fig:workflow} provides an overview of the workflow of tools used in our evaluation.

\subsection{Modeling Area Using JJ-Complexity}
\label{sec:jj_modeling}

The area of a superconducting circuit is denoted by a term, called as {\em JJ-complexity}. The JJ-complexity is the number of JJs required to design a logic block. A logic block consists of logic gates and Josephson Junction Transmission Lines(JTLs). As JJ-based logic gates have limited driving strength, JTLs are inserted to facilitate the desired fanout. In this paper, we use JJ-complexity as a key figure of merit, similar to prior superconducting system designs~\cite{dorojevets2015towards,dorojevets2015fast}. We evaluate the system level JJ-complexity by computing gate JJ-complexity and interconnect JJ-complexity.

\vspace{0.05 in}
\noindent{\bf Gate JJ-Complexity:}  To evaluate gate JJ-complexity, we use the RQL standard cell library and \texttt{Yosys}~\cite{wolf2016yosys,oberg2011superconducting}, an open-source synthesis tool. \texttt{Yosys} enables us to derive the gate level netlist using only RQL standard cells. \texttt{Yosys} uses \texttt{ABC}~\cite{mishchenko2007abc}, that allows it to map a design's gate level representation to a target custom library (which is the RQL cell library for our analysis). We process the netlist to compute the gate JJ-complexity by determining the number of gates used of each type. Additionally, we also use \texttt{Synopsys Design Compiler}, a standard CMOS synthesis tool to generate the gate-level design and post-process the design to optimize it using RQL specific standard cells. The gate JJ-complexity obtained is the same from both techniques. Note that the lack of place and route tools, and restricted access to foundry models forces the superconducting logic designers to use manual routing to compute JJ-complexity.  

\begin{figure}[!htb]
\centering

    \includegraphics[width=1.0\columnwidth]{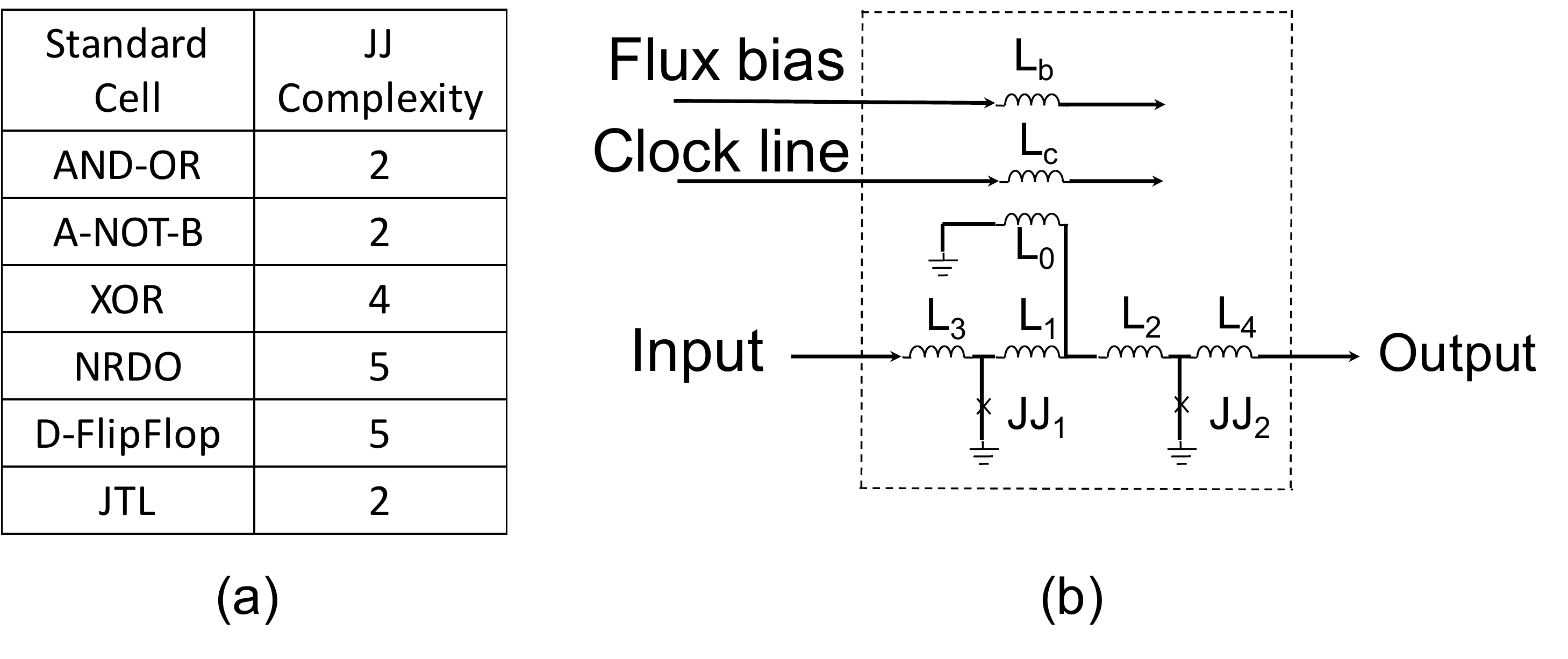}
    \caption{(a) JJ Complexity of RQL Standard Cells (b) Josephson Transmission Line (JTL)}
\label{fig:JTL}    
\end{figure}


\noindent{\bf Interconnect JJ Complexity:}  As RQL gates have limited driving strength, JTLs are used to drive gates. As shown in Figure~\ref{fig:JTL}(b), each JTL comprises 2 JJs. 
JTLs enable fan-out capacity similar to buffers in traditional CMOS circuits and limit clock-skew and jitter. Due to limited driving strength, RQL gates require one JTL for every output load. We process the \texttt{Yosys} generated netlist and determine the fanout for every input and output port and internal wires. To account for JTL overheads we compute the number of JJs required using the rules based on ~\cite{herr2011ultra}: 
\begin{enumerate}
    \item A JTL is added after a series of five logic gates to suppress clock skew and jitter.
    \item A JTL is required per fanout (a gate can drive a JTL, and a JTL can drive a gate and a JTL).
    \item XOR gates need extra JTL because they operate at the phase boundary (RQL uses a four phase clock, two clock lines with a phase difference of $\pi/2$ provide two phases each~\cite{oberg2011superconducting}).
\end{enumerate}

Our analysis (Figure~\ref{fig:fanoutdis}) shows that most of the gates drive either 1 or 2 gates, and the percentage of gates that drive more than 2 gates is quite small (less than 1\%). Given that approximately half the gates drive exactly 2 gates, the overhead of additional JTL due to fanout is approximately 50\% for our baseline implementation. 

\vspace{0.05 in}
\noindent{\bf System JJ complexity:}  Full system design using superconducting technology requires JJs for implementing logic and enabling signal routing and fanout. We derive the system JJ-complexity $(JJ_{system})$ by adding gate JJ-complexity $(JJ_{gate})$ and interconnect JJ-complexity $(JJ_{interconnect})$ as shown in Equation 1.

\begin{equation}
{JJ_{system} = JJ_{gate} + JJ_{interconnect} }
\end{equation}

Table~\ref{tab:jjcomplex_b} shows the JJ-complexity of some commonly used logic blocks. For validation, we compare our method of evaluating JJ-complexity against published designs that use foundry RQL standard cell library based on foundry models and observe that our estimates are within 12\% of the numbers reported in prior work ~\cite{dorojevets2015towards,dorojevets2015fast}.

\begin{table}[hb!]
\centering
\begin{small}
\caption{Evaluations for proposed Design Methodology}
\setlength{\tabcolsep}{0.05cm} 
\renewcommand{\arraystretch}{1.2}
\label{tab:contribution}
\begin{tabular}{ |c|c|c|c| } 
\hline
	 Logic  & Estimated  & Reported JJ & Percentage\\ 
	 block &   JJ Complexity & Complexity~\cite{dorojevets2015towards}          & Error                         \\ \hline \hline

    \texttt{32 bit RCA Adder} &  1316 &  1410 & 6.6 \% \\ \hline
    \texttt{32 bit KSA Adder} & 3992 & 4160 & 4.0 \% \\ \hline 
    \texttt{Integer Multiplier } & 33320 &  37782 & 11.8 \%\\ \hline 
                
\end{tabular}
\label{tab:jjcomplex_b}
\end{small}
\vspace{-0.15 in}
\end{table}

\ignore{

\begin{table}[!htb]
\vspace{-0.1in}
    \caption{JJ Complexity and JTL for gates and logic circuits (RCA, Ripple-Carry and KSA, Kogge-Stone)}
   \vspace{-0.05in}
    \begin{subtable}{.35\linewidth}
      \begin{small}
\begin{center}

\setlength{\tabcolsep}{0.05cm} 
\begin{tabular}{ | c | c | c | c | c | c |}

\hline
	Logic  &   JJ        \\ 
	Gate   &   Complexity                                         \\ \hline \hline

	\texttt{OR}  & 2     \\ \hline
    \texttt{AND}  & 2   \\ \hline
	\texttt{XOR}    & 4    \\ \hline 
    \texttt{NDRO}   & 5    \\ \hline
    \texttt{NOT}    & 4    \\ \hline 
    \texttt{D-flop}  & 12    \\ \hline 

\end{tabular}
\end{center}
\end{small}
    \end{subtable}%
    \begin{subtable}{.5\linewidth}
      \begin{small}
\begin{center}
\setlength{\tabcolsep}{0.05cm} 
\begin{tabular}{ | c | c | c | c | c | c |}

\hline
	 Logic  & JJ  &JTL   \\ 
	 block &   Complexity & overhead                                      \\ \hline \hline

	\texttt{2:1 Mux} &  18  & 30\% \\ \hline
    \texttt{8:1 Mux} &  56 & 35\%\\ \hline
	\texttt{8 bit RCA Adder}  & 320 & 55\%  \\ \hline 
    \texttt{32 bit RCA Adder} &  1280 &  55\%  \\ \hline
    \texttt{8 bit KSA Adder} & 1732 & 75.3\%   \\ \hline 
    \texttt{32 bit KSA Adder} & 4308 &  76\%   \\ \hline 

\end{tabular}
\label{table:jjcomplex}
\end{center}
\end{small}
    \end{subtable}
    
\label{tab:jjcomplex}
\end{table}

}

\subsection{Modeling Power}
RQL delivers power to on-chip devices through inductive coupling to an AC transmission line. As a result, RQL circuits dissipate negligible static power. RQL uses reciprocal data encoding where ``0" is represented by the absence of SFQ pulses. Therefore, the dynamic power dissipation in RQL circuits result from only digital ``1"s, and digital ``0"s do not dissipate power. The total power dissipated (P$_{dynamic}$) by an RQL circuit is  given by Equation 2.

\begin{equation}
{P_{dynamic} = \frac{2}{3} \cdot n \cdot f \cdot I_c \cdot \phi_0 \cdot \alpha }
\end{equation}

where, \textit{n} is the number of JJs, \textit{f} is the frequency, I$_c$ is the critical current, $\phi_0$ is a universal constant, and $\alpha$ is the activity factor (or the percentage of JJs switching to ``1" state). The power dissipated by the superconducting logic is directly proportional to the critical current which depends on the device fabrication technology and foundry services. For our evaluations, we assume the critical current to be $38 \mu A$. However, a conservative analysis of critical current reveals that it is possible to reduce it to $10 \mu A$ without substantial impact on the bit error rate~\cite{herr2011ultra}. We determine the activity factor ($\alpha$) of a design by counting the number of ``1"s from the value change dump (VCD) file of random simulations. 
We evaluate the total power consumption ($P_{total}$) by multiplying the power dissipated by the design at 4.2K with a cooling overhead as shown in Equation 3.

\begin{equation}
{P_{total} = cooling_{factor} \cdot P_{dynamic} }
\end{equation}

\vspace{.05in}

\begin{tcolorbox}[top=3pt,bottom=3pt]
{\bf Note:} For all accelerators implemented using the JJ technology, we include a cooling overhead of 300x in our energy-efficiency calculations. 
\end{tcolorbox}

\subsection{Modeling Performance}

In order to model performance, we count the number of JJs in the critical path of a design and multiply it by the switching time of a JJ. We assume a uniform JJ switching time of 2 ps~\cite{herr2011ultra} for all devices. Switching time can be improved by using larger feature size. However, for our analysis, we lack the design and layout tools to study such optimizations.

\ignore{

\begin{table}[hb!]
\label{table:3}
\begin{small}
\caption{Performance of RQL hash engines}
\setlength{\tabcolsep}{0.05cm} 
\renewcommand{\arraystretch}{1.2}
\label{tab:jj_nums_basic}
\begin{tabular}{ |c|c| } 
\hline
Block                               & JJ-Complexity     \\
                                        \hline 
                                        \hline
2:1 MUX                             &  18               \\
\hline
1-bit half adder                    &                   \\
\hline
1-bit full adder                    &  40               \\
\hline
Majority Logic                      &                   \\
\hline
Choose Function                     &                   \\
\hline
Sigma Function                      &                   \\
\hline
Sum Function                        &                   \\
\hline
\end{tabular}
\label{table:jj_nums_basic}
\end{small}
\end{table}

}

\section{Related Work}
\textbf{Superconducting circuits:} A Josephson junction based processor was proposed as early as 1980 ~\cite{anacker1980josephson}. A number of circuits were demonstrated in RSFQ logic in the 1990s, including DSPs ~\cite{gupta2007digital}, microprocessor components ~\cite{mukhanov1993rapid, mukhanov1993rsfq, filippov20118, dorojevets20138, yamanashi2007design}, mixed signal devices ~\cite{rylov1995superconducting, inamdar2009progress, kirichenko2003multi}, floating point units ~\cite{mukhanov1995implementation, kainuma2011design}. However, due to static power dissipation challenges and high device counts per logic gate, RSFQ circuits faced scalability issues. With the introduction of the RQL family of logic gates, designers were able to mitigate the disadvantages of RSFQ. 
So far, several  RQL family circuits are demonstrated including shift registers, an 8-bit carry-lookahead adder, shift register yield vehicles, transmission driver systems, and serial data receiver systems ~\cite{herr2011ultra, herr20138, herr2015reproducible,herr2018superconducting,shauck2018reciprocal}. Dorojevets et. al. present resource estimates for 32-bit and 64-bit integer and floating-point ALU, on-chip storage elements, and bloom filters ~\cite{dorojevets2015towards, dorojevets2015fast,dorojevets2018energy}. Holmes et al. analyze the feasibility of HPC systems with 1000 PFLOPs~\cite{holmes2013energy}.

\textbf{SHA Designs:} SHA-256 optimizations include changing the computational platform from general purpose processors to FPGAs~\cite{chaves2008cost, chaves2006improving, esuruoso2011high, ahmad2005hardware, mcevoy2006optimisation} and ASICs~\cite{dadda2004asic, dadda2004design, kim2008efficient, satoh2007asic}. Hardware optimizations of SHA engines include use of carry save adders and combination of different types of adders~\cite{dadda2004asic, dadda2004design, lien20041, mcevoy2006optimisation}, pipeline designs~\cite{macchetti2005quasi, crowe2004single, lien20041, macchetti2005quasi, mcevoy2006optimisation}, delay balancing~\cite{dadda2004design}, operation rescheduling~\cite{dadda2004design, ting2002fpga, chaves2008cost, chaves2006improving}.

\textbf{Reliable SHA cores:} Prior fault tolerant schemes for SHA hardware use triple modular redundancy ~\cite{juliato2009efficient}, register protection using Hamming codes~\cite{juliato2008seu}, and SHA cores with inbuilt self-checking mechanisms~\cite{michail2015hardware, michail2016design}. These schemes assume uncorrelated errors and incur significant area and complexity.

\section{Conclusion}


In this paper, we evaluate the system level performance and energy improvements for an accelerator built with Josephson junction technology. We focus on three JJ-technology challenges: low device density, limited fanout, and correlated faults due to flux trapping. To leverage the existing technology with limited device density, we focus on SHA-256 engines, that are commonly used in bitcoin-mining accelerators. This application has high computational intensity, tiny memory footprint, and energy-efficiency is a key metric.  

A direct translation of CMOS design to JJ design of a baseline~\cite{Goldstrike} provides 10x improvement in energy-efficiency (GH/J).  We highlight the fan-out overhead in JJ technology, and how it impacts the design choices for arithmetic units and pipeline design. We study a technology-aware design that improves the performance by 1.6x while boosting the energy efficiency to 12x over CMOS baseline. We present a unique reliability challenge in JJ technology and propose a fault-tolerant design that can protect against large granularity faults that occur due to this new failure mode. Moreover, we utilize this fault-tolerant design to enable better than worse case design that enables scaling of the critical current without sacrificing functionality and providing a 46x improvement in energy efficiency over CMOS design.  We introduce a methodology for estimating area, performance, and power of accelerators built in superconducting logic. Such a workflow can help other researchers in exploring designs using  this technology.  While we evaluate SHA-256 as an example, the JJ technology is also applicable to other domains.


\begin{acks}
We thank Srilatha Manne, Elnaz Ansari, Zachary Myers for the technical discussions and feedback. This work was supported by a gift from  Microsoft Research.
\end{acks}

\bibliographystyle{ieeetr}
\bibliography{ref}

\begin{thebibliography}{10}

\bibitem{TPU}
N.~P. Jouppi, C.~Young, N.~Patil, D.~Patterson, G.~Agrawal, R.~Bajwa, S.~Bates,
  S.~Bhatia, N.~Boden, A.~Borchers, {\em et~al.}, ``In-datacenter performance
  analysis of a tensor processing unit,'' in {\em Computer Architecture (ISCA),
  2017 ACM/IEEE 44th Annual International Symposium on}, pp.~1--12, IEEE, 2017.

\bibitem{brainwave}
J.~Fowers, K.~Ovtcharov, M.~Papamichael, T.~Massengill, M.~Liu, D.~Lo,
  S.~Alkalay, M.~Haselman, L.~Adams, M.~Ghandi, {\em et~al.}, ``A configurable
  cloud-scale dnn processor for real-time ai,'' in {\em 2018 ACM/IEEE 45th
  Annual International Symposium on Computer Architecture (ISCA)}, IEEE, 2018.

\bibitem{herr2011ultra}
Q.~P. Herr, A.~Y. Herr, O.~T. Oberg, and A.~G. Ioannidis, ``Ultra-low-power
  superconductor logic,'' {\em Journal of applied physics}, vol.~109, no.~10,
  p.~103903, 2011.

\bibitem{dorojevets2015towards}
M.~Dorojevets, Z.~Chen, C.~L. Ayala, and A.~K. Kasperek, ``Towards 32-bit
  energy-efficient superconductor rql processors: the cell-level design and
  analysis of key processing and on-chip storage units,'' {\em IEEE
  Transactions on Applied Superconductivity}, vol.~25, no.~3, pp.~1--8, 2015.

\bibitem{pedram}
A.~Pedram, S.~Richardson, M.~Horowitz, S.~Galal, and S.~Kvatinsky, ``Dark
  memory and accelerator-rich system optimization in the dark silicon era,''
  {\em IEEE Design \& Test}, vol.~34, no.~2, pp.~39--50, 2017.

\bibitem{holmes2013energy}
D.~S. Holmes, A.~L. Ripple, and M.~A. Manheimer, ``Energy efficient
  superconducting computing power budgets and requirements,'' {\em IEEE
  Transactions on Applied Superconductivity}, vol.~23, no.~3,
  pp.~1701610--1701610, 2013.

\bibitem{mit_ll}
M.~L. Laboratory, ``Beyond cmos superconducting digital circuits,''

\bibitem{tolpygo2016superconductor}
S.~K. Tolpygo, ``Superconductor digital electronics: Scalability and energy
  efficiency issues,'' {\em Low Temperature Physics}, vol.~42, no.~5,
  pp.~361--379, 2016.

\bibitem{Goldstrike}
J.~Barkatullah and T.~Hanke, ``Goldstrike 1: Cointerra's first-generation
  cryptocurrency mining processor for bitcoin,'' {\em IEEE micro}, vol.~35,
  no.~2, pp.~68--76, 2015.

\bibitem{debitcoin}
A.~de~Vries, ``Bitcoin's growing energy problem,'' {\em Joule}, vol.~2, no.~5,
  pp.~801--805, 2018.

\bibitem{oberg2011superconducting}
O.~T. Oberg, {\em Superconducting logic circuits operating with reciprocal
  magnetic flux quanta}.
\newblock University of Maryland, College Park, 2011.

\bibitem{herr20138}
A.~Y. Herr, Q.~P. Herr, O.~T. Oberg, O.~Naaman, J.~X. Przybysz, P.~Borodulin,
  and S.~B. Shauck, ``An 8-bit carry look-ahead adder with 150 ps latency and
  sub-microwatt power dissipation at 10 ghz,'' {\em Journal of Applied
  Physics}, vol.~113, no.~3, p.~033911, 2013.

\bibitem{herr2015reproducible}
Q.~P. Herr, J.~Osborne, M.~J. Stoutimore, H.~Hearne, R.~Selig, J.~Vogel,
  E.~Min, V.~V. Talanov, and A.~Y. Herr, ``Reproducible operating margins on a
  72 800-device digital superconducting chip,'' {\em Superconductor Science and
  Technology}, vol.~28, no.~12, p.~124003, 2015.

\bibitem{herr2018superconducting}
Q.~P. Herr, E.~Rudman, J.~D. Egan, and V.~V. Talanov, ``Superconducting
  transmission driver system,'' May~24 2018.
\newblock US Patent App. 15/356,049.

\bibitem{shauck2018reciprocal}
S.~B. SHAUCK, ``Reciprocal quantum logic (rql) serial data receiver system,''
  Sept.~25 2018.
\newblock US Patent App. 10/083,148.

\bibitem{dorojevets2015fast}
M.~Dorojevets and Z.~Chen, ``Fast pipelined storage for high-performance
  energy-efficient computing with superconductor technology,'' in {\em Emerging
  Technologies for a Smarter World (CEWIT), 2015 12th International Conference
  \& Expo on}, pp.~1--6, IEEE, 2015.

\bibitem{dorojevets2018energy}
M.~Dorojevets, ``Energy-efficient superconductor bloom filters for streaming
  data inspection,'' {\em IEEE Transactions on Dependable and Secure
  Computing}, 2018.

\bibitem{wright2018standards}
F.~Wright and T.~M. Conte, ``Standards: Roadmapping computer technology trends
  enlightens industry,'' {\em Computer}, no.~6, pp.~100--103, 2018.

\bibitem{esmaeilzadeh2011dark}
H.~Esmaeilzadeh, E.~Blem, R.~S. Amant, K.~Sankaralingam, and D.~Burger, ``Dark
  silicon and the end of multicore scaling,'' in {\em Computer Architecture
  (ISCA), 2011 38th Annual International Symposium on}, pp.~365--376, IEEE,
  2011.

\bibitem{nakamoto}
S.~Nakamoto, ``Bitcoin: A peer-to-peer electronic cash system,''

\bibitem{o2014bitcoin}
K.~J. O'Dwyer and D.~Malone, ``Bitcoin mining and its energy footprint,'' 2014.

\bibitem{magaki2016asic}
I.~Magaki, M.~Khazraee, L.~V. Gutierrez, and M.~B. Taylor, ``Asic clouds:
  specializing the datacenter,'' in {\em ACM SIGARCH Computer Architecture
  News}, vol.~44, pp.~178--190, IEEE Press, 2016.

\bibitem{taylor2017evolution}
M.~B. Taylor, ``The evolution of bitcoin hardware,'' {\em Computer}, vol.~50,
  no.~9, pp.~58--66, 2017.

\bibitem{dorojevets20138}
M.~Dorojevets, C.~L. Ayala, N.~Yoshikawa, and A.~Fujimaki, ``8-bit asynchronous
  sparse-tree superconductor rsfq arithmetic-logic unit with a rich set of
  operations,'' {\em IEEE Transactions on Applied Superconductivity}, vol.~23,
  no.~3, pp.~1700104--1700104, 2013.

\bibitem{ting2002fpga}
K.~K. Ting, S.~C. Yuen, K.-H. Lee, and P.~H. Leong, ``An fpga based sha-256
  processor,'' in {\em International Conference on Field Programmable Logic and
  Applications}, pp.~577--585, Springer, 2002.

\bibitem{dadda2004design}
L.~Dadda, M.~Macchetti, and J.~Owen, ``The design of a high speed asic unit for
  the hash function sha-256 (384, 512),'' in {\em Proceedings of the conference
  on Design, automation and test in Europe-Volume 3}, p.~30070, IEEE Computer
  Society, 2004.

\bibitem{dadda2004asic}
L.~Dadda, M.~Macchetti, and J.~Owen, ``An asic design for a high speed
  implementation of the hash function sha-256 (384, 512),'' in {\em Proceedings
  of the 14th ACM Great Lakes symposium on VLSI}, pp.~421--425, ACM, 2004.

\bibitem{eckert1953memory}
J.~J.~P. Eckert and J.~W. Mauchly, ``Memory system,'' Feb.~24 1953.
\newblock US Patent 2,629,827.

\bibitem{auerbach1949mercury}
I.~L. Auerbach, J.~P. Eckert, R.~Shaw, and C.~Sheppard, ``Mercury delay line
  memory using a pulse rate of several megacycles,'' {\em Proceedings of the
  IRE}, vol.~37, no.~8, pp.~855--861, 1949.

\bibitem{vilim2016approximate}
M.~Vilim, H.~Duwe, and R.~Kumar, ``Approximate bitcoin mining,'' in {\em Design
  Automation Conference (DAC), 2016 53nd ACM/EDAC/IEEE}, pp.~1--6, IEEE, 2016.

\bibitem{holmes2017non}
D.~S. Holmes and J.~McHenry, ``Non-normal critical current distributions in
  josephson junctions with aluminum oxide barriers,'' {\em IEEE Transactions on
  Applied Superconductivity}, vol.~27, no.~4, pp.~1--5, 2017.

\bibitem{wolf2016yosys}
C.~Wolf, ``Yosys open synthesis suite,'' 2016.

\bibitem{mishchenko2007abc}
A.~Mishchenko {\em et~al.}, ``Abc: A system for sequential synthesis and
  verification,'' {\em URL http://www. eecs. berkeley. edu/alanmi/abc}, p.~17,
  2007.

\bibitem{anacker1980josephson}
W.~Anacker, ``Josephson computer technology: An ibm research project,'' {\em
  IBM Journal of research and development}, vol.~24, no.~2, pp.~107--112, 1980.

\bibitem{gupta2007digital}
D.~Gupta, T.~V. Filippov, A.~F. Kirichenko, D.~E. Kirichenko, I.~V. Vernik,
  A.~Sahu, S.~Sarwana, P.~Shevchenko, A.~Talalaevskii, and O.~A. Mukhanov,
  ``Digital channelizing radio frequency receiver,'' {\em IEEE Transactions on
  applied superconductivity}, vol.~17, no.~2, pp.~430--437, 2007.

\bibitem{mukhanov1993rapid}
O.~A. Mukhanov, ``Rapid single flux quantum (rsfq) shift register family,''
  {\em IEEE transactions on applied superconductivity}, vol.~3, no.~1,
  pp.~2578--2581, 1993.

\bibitem{mukhanov1993rsfq}
O.~A. Mukhanov, ``Rsfq 1024-bit shift register for acquisition memory,'' {\em
  IEEE transactions on applied superconductivity}, vol.~3, no.~4,
  pp.~3102--3113, 1993.

\bibitem{filippov20118}
T.~Filippov, M.~Dorojevets, A.~Sahu, A.~Kirichenko, C.~Ayala, and O.~Mukhanov,
  ``8-bit asynchronous wave-pipelined rsfq arithmetic-logic unit,'' {\em IEEE
  Transactions on Applied Superconductivity}, vol.~21, no.~3, pp.~847--851,
  2011.

\bibitem{yamanashi2007design}
Y.~Yamanashi, M.~Tanaka, A.~Akimoto, H.~Park, Y.~Kamiya, N.~Irie, N.~Yoshikawa,
  A.~Fujimaki, H.~Terai, and Y.~Hashimoto, ``Design and implementation of a
  pipelined bit-serial sfq microprocessor, core 1$\beta$,'' {\em IEEE
  transactions on applied superconductivity}, vol.~17, no.~2, pp.~474--477,
  2007.

\bibitem{rylov1995superconducting}
S.~V. Rylov and R.~P. Robertazzi, ``Superconducting high-resolution a/d
  converter based on phase modulation and multichannel timing arbitration,''
  {\em IEEE Transactions on Applied Superconductivity}, vol.~5, no.~2,
  pp.~2260--2263, 1995.

\bibitem{inamdar2009progress}
A.~Inamdar, S.~Rylov, A.~Talalaevskii, A.~Sahu, S.~Sarwana, D.~E. Kirichenko,
  I.~V. Vernik, T.~V. Filippov, and D.~Gupta, ``Progress in design of improved
  high dynamic range analog-to-digital converters,'' {\em IEEE Transactions on
  Applied Superconductivity}, vol.~19, no.~3, pp.~670--675, 2009.

\bibitem{kirichenko2003multi}
A.~Kirichenko, S.~Sarwana, D.~Gupta, I.~Rochwarger, and O.~Mukhanov,
  ``Multi-channel time digitizing systems,'' {\em IEEE transactions on applied
  superconductivity}, vol.~13, no.~2, pp.~454--458, 2003.

\bibitem{mukhanov1995implementation}
O.~A. Mukhanov and A.~F. Kirichenko, ``Implementation of a fft radix 2
  butterfly using serial rsfq multiplier-adders,'' {\em IEEE Transactions on
  Applied Superconductivity}, vol.~5, no.~2, pp.~2461--2464, 1995.

\bibitem{kainuma2011design}
T.~Kainuma, Y.~Shimamura, F.~Miyaoka, Y.~Yamanashi, N.~Yoshikawa, A.~Fujimaki,
  K.~Takagi, N.~Takagi, and S.~Nagasawa, ``Design and implementation of
  component circuits of an sfq half-precision floating-point adder using
  10-ka/cm nb process,'' {\em IEEE Transactions on Applied Superconductivity},
  vol.~21, no.~3, pp.~827--830, 2011.

\bibitem{chaves2008cost}
R.~Chaves, G.~Kuzmanov, L.~Sousa, and S.~Vassiliadis, ``Cost-efficient sha
  hardware accelerators,'' {\em IEEE transactions on very large scale
  integration (VLSI) Systems}, vol.~16, no.~8, pp.~999--1008, 2008.

\bibitem{chaves2006improving}
R.~Chaves, G.~Kuzmanov, L.~Sousa, and S.~Vassiliadis, ``Improving sha-2
  hardware implementations,'' in {\em International Workshop on Cryptographic
  Hardware and Embedded Systems}, pp.~298--310, Springer, 2006.

\bibitem{esuruoso2011high}
O.~Esuruoso, ``High speed fpga implementation of cryptographic hash function,''
  2011.

\bibitem{ahmad2005hardware}
I.~Ahmad and A.~S. Das, ``Hardware implementation analysis of sha-256 and
  sha-512 algorithms on fpgas,'' {\em Computers \& Electrical Engineering},
  vol.~31, no.~6, pp.~345--360, 2005.

\bibitem{mcevoy2006optimisation}
R.~P. McEvoy, F.~M. Crowe, C.~C. Murphy, and W.~P. Marnane, ``Optimisation of
  the sha-2 family of hash functions on fpgas,'' in {\em Emerging VLSI
  Technologies and Architectures, 2006. IEEE Computer Society Annual Symposium
  on}, pp.~6--pp, IEEE, 2006.

\bibitem{kim2008efficient}
M.~Kim, J.~Ryou, and S.~Jun, ``Efficient hardware architecture of sha-256
  algorithm for trusted mobile computing,'' in {\em International Conference on
  Information Security and Cryptology}, pp.~240--252, Springer, 2008.

\bibitem{satoh2007asic}
A.~Satoh and T.~Inoue, ``Asic-hardware-focused comparison for hash functions
  md5, ripemd-160, and shs,'' {\em INTEGRATION, the VLSI journal}, vol.~40,
  no.~1, pp.~3--10, 2007.

\bibitem{lien20041}
R.~Lien, T.~Grembowski, and K.~Gaj, ``A 1 gbit/s partially unrolled
  architecture of hash functions sha-1 and sha-512,'' in {\em Cryptographers’
  Track at the RSA Conference}, pp.~324--338, Springer, 2004.

\bibitem{macchetti2005quasi}
M.~Macchetti and L.~Dadda, ``Quasi-pipelined hash circuits,'' in {\em Computer
  Arithmetic, 2005. ARITH-17 2005. 17th IEEE Symposium on}, pp.~222--229, IEEE,
  2005.

\bibitem{crowe2004single}
F.~Crowe, A.~Daly, T.~Kerins, and W.~Marnane, ``Single-chip fpga implementation
  of a cryptographic co-processor,'' in {\em Field-Programmable Technology,
  2004. Proceedings. 2004 IEEE International Conference on}, pp.~279--285,
  IEEE, 2004.

\bibitem{juliato2009efficient}
M.~Juliato, C.~Gebotys, and R.~Elbaz, ``Efficient fault tolerant sha-2 hash
  functions for space applications,'' in {\em Aerospace conference, 2009 IEEE},
  pp.~1--16, IEEE, 2009.

\bibitem{juliato2008seu}
M.~Juliato and C.~Gebotys, ``Seu-resistant sha-256 design for security in
  satellites,'' in {\em Signal Processing for Space Communications, 2008. SPSC
  2008. 10th International Workshop on}, pp.~1--7, IEEE, 2008.

\bibitem{michail2015hardware}
H.~E. Michail, A.~Kotsiolis, A.~Kakarountas, G.~Athanasiou, and C.~Goutis,
  ``Hardware implementation of the totally self-checking sha-256 hash core,''
  in {\em EUROCON 2015-International Conference on Computer as a Tool
  (EUROCON), IEEE}, pp.~1--5, IEEE, 2015.

\bibitem{michail2016design}
H.~E. Michail, G.~S. Athanasiou, G.~Theodoridis, A.~Gregoriades, and C.~E.
  Goutis, ``Design and implementation of totally-self checking sha-1 and
  sha-256 hash functions’ architectures,'' {\em Microprocessors and
  Microsystems}, vol.~45, pp.~227--240, 2016.

\end{thebibliography}

\end{document}